\documentclass[traditabstract]{aa}

\usepackage{amsmath}
\usepackage{natbib}
\usepackage{graphicx}
\usepackage{txfonts}
\usepackage{hyperref}

\newcommand{\rxte}{\textit{RXTE}}
\newcommand{\inte}{\textit{INTEGRAL}}
\newcommand{\Efold}{\ensuremath{E_{\mathrm{fold}}}}

\newcommand{\chisq}{\ensuremath{\chi^2_{\mathrm{red}}}}

\graphicspath{{./pics/}}

\begin{document}

\title{No anti-correlation between cyclotron line energy and X-ray
  flux in 4U~0115+634}

\author{
 Sebastian M\"uller \inst{1}
 \and Carlo Ferrigno \inst{2}
 \and Matthias K\"uhnel \inst{1}
 \and Gabriele Sch\"onherr \inst{3}
 \and Peter A. Becker \inst{4}
 \and Michael T. Wolff \inst{5}
 \and Dominik Hertel \inst{1}
 \and Fritz-Walter Schwarm \inst{1}
 \and Victoria Grinberg \inst{1}
 \and Maria Obst \inst{1}
 \and Isabel Caballero \inst{6}
 \and Katja Pottschmidt \inst{7}
 \and Felix F\"urst \inst{8}
 \and Ingo Kreykenbohm \inst{1}
 \and Richard E. Rothschild \inst{9}
 \and Paul Hemphill \inst{9}
 \and Silvia Mart\'inez N\'u\~nez \inst{10}
 \and \mbox{Jos\'e M. Torrej\'on} \inst{10}
 \and Dmitry Klochkov \inst{11}
 \and R\"udiger Staubert \inst{11}
 \and J\"orn Wilms \inst{1}
}

\institute{
      Dr. Karl Remeis-Observatory \& ECAP, Universit\"at 
      Erlangen-N\"urnberg, Sternwartstr.~7, 96049 Bamberg, Germany\\
      email: \texttt{Sebastian.Mueller@sternwarte.uni-erlangen.de}
 \and ISDC Data Center for Astrophysics, University of Geneva, 16
      Chemin d'\'Ecogia, 1290 Versoix, Switzerland 
 \and Leibniz-Institut f\"ur Astrophysik Potsdam, An der Sternwarte
      16, 14482 Potsdam, Germany
 \and School of Physics, Astronomy, and Computational Sciences, MS
      5C3, George Mason University,
      4400 University Drive, Fairfax, VA 22030, USA
 \and Space Science Division, Naval Research Laboratory, Code 7655,
      4555 Overlook Ave., S.W., Washington, DC 20375, USA
 \and CEA Saclay, DSM/IRFU/SAp-UMR AIM (7158) CNRS/CEA/Universit\'e
      Paris 7, Diderot, 91191 Gif sur Yvette, France 
 \and CRESST and NASA Goddard Space Flight Center, Greenbelt, MD 20771, USA, and
      Center for Space Science and Technology, UMBC, Baltimore, MD 21250,
      USA 
 \and Space Radiation Laboratory, 290-17 Cahill, Caltech, 1200 East
      California Blvd., Pasadena, CA 91125, USA
 \and Center for Astrophysics and Space Sciences, University of 
      California, San Diego, La Jolla, CA 92093, USA
 \and Instituto Universitario de F\'isica Aplicada a las Ciencias y las 
      Tecnolog\'ias, University of Alicante, P.O.\ Box 99, 03080
      Alicante, Spain 
 \and Institut f\"ur Astronomie und Astrophysik, Universit\"at
      T\"ubingen, Sand 1, 72076 T\"ubingen, Germany
}

\abstract{We report on an outburst of the high mass X-ray binary
  4U~0115+63 with a pulse period of 3.6\,s in 2008 March/April as
  observed with \rxte\ and \inte. During the outburst the neutron
  star's luminosity varied by a factor of 10 in the 3--50\,keV band.
  In agreement with earlier work we find evidence for five cyclotron
  resonance scattering features at $\sim$10.7, 21.8, 35.5, 46.7, and
  59.7\,keV. Previous work had found an anti-correlation between the
  fundamental cyclotron line energy and the X-ray flux. We show that
  this apparent anti-correlation is probably due to the unphysical
  interplay of parameters of the cyclotron line with the continuum
  models used previously, e.g., the negative and positive exponent
  power law (NPEX). For this model, we show that cyclotron line
  modeling erroneously leads to describing part of the exponential
  cutoff and the continuum variability, and not the cyclotron
  lines. When the X-ray continuum is modeled with a simple
  exponentially cutoff power law modified by a Gaussian emission
  feature around 10\,keV, the correlation between the line energy and
  the flux vanishes and the line parameters remain virtually constant
  over the outburst. We therefore conclude that the previously
  reported anti-correlation is an artifact of the assumptions adopted
  in the modeling of the continuum.}

\date{Received DATE  / Accepted DATE}
\keywords{X-rays: binaries - 
pulsars: individual \object{4U~0115+634} -
Magnetic Fields 
}

\maketitle
\section{Introduction}

In all accreting X-ray pulsars the accretion geometry is strongly
affected by the magnetic field of the neutron star, which has values
of the order of a few $10^{12}$\,G at the magnetic poles. The accreted
matter is forced to follow the magnetic field lines from the Alfv\'en
radius inward to the accretion columns, located at the magnetic poles
of the neutron star. The shape of these columns remains an open
question. The simplest models assume cylinder or slab-like geometries,
although more complicated shapes have been discussed
\citep{Meszaros1984a,Becker2012a}. On each magnetic pole of the
neutron star a hot spot is created, emitting thermal photons into the
accretion column \citep{Davidson1973a}. These photons are upscattered
in energy by inverse Compton scattering within the accretion column,
forming the hard X-ray spectrum \citep{Becker2007a,Ferrigno2009a}.
Depending on the source luminosity, and thus mass accretion rate,
$\dot{M}$, and deceleration mechanism, a radiation dominated shock can
form above the surface of the neutron star. See \citet{Becker2012a}
for a discussion of the different modes of accretion in the accretion
column.

Due to the strong magnetic field of accreting X-ray pulsars, cyclotron
resonance scattering features (cyclotron lines; CRSFs) are
observable in many X-ray pulsars. These features have
been reported for about 20 X-ray pulsars \citep{Caballero2011a}. CRSFs
are produced by photons generated in the accretion column of the
neutron star interacting with electrons \citep[][and references
therein]{Schoenherr2007a,Meszaros1992a}. The motion of these electrons
perpendicular to the $B$-field is quantized into Landau levels with
energy differences of about
\begin{equation}\label{eq:12B12}
\Delta E\approx 12\,\mathrm{keV}\left(\frac{B}{10^{12}\,\mathrm{G}}\right).
\end{equation}
Photons inelastically scattering off these electrons will imprint
absorption-line like features onto the spectral continuum.

Measuring the energy of the CRSF yields information about the
$B$-field strength in the line forming region. Since to first order
the $B$-field of a neutron star can be assumed to be dipole-like, in
many models the centroid energy of CRSFs depends on the characteristic
emission height, i.e., the altitude of the line forming region
\citep{Basko1976a,Burnard1991a,Mihara2004a}. Since this altitude
depends on the complex relation between the deceleration of the
accreted material and the mass accretion rate, it is expected to
depend on the source luminosity. Such a dependence has indeed been
observed. Three types of behaviour have been observed. One type
shows a positive correlation between the X-ray luminosity
$L_\mathrm{X}$ and the cyclotron line energy $E_\mathrm{cyc}$
\citep[e.g., \object{Her~X-1},][]{Staubert2007a}. The other type
shows a negative correlation between $L_\mathrm{X}$ and
$E_\mathrm{cyc}$ \citep[e.g., \object{V0332+53},][and references
therein]{Nakajima2010a}. Finally, there are also cyclotron sources
known with a constant cyclotron line energy \citep[e.g.,
\object{A0535+26},][]{Caballero2007a}.

As suggested by \citet{Staubert2007a} and \citet{Klochkov2011a} and
then quantified by \citet{Becker2012a}, these different relations
between $L_\mathrm{X}$ and $E_\mathrm{cyc}$ could be due to the
different deceleration mechanisms for the accreted material. Above a
certain critical luminosity, $L_\mathrm{crit}$, the accreted material
is decelerated mainly by the radiation pressure. In this situation, an
increase in luminosity stops the material earlier, leading to the line
forming region being in a region of lower $B$-field strength and
yielding a negative correlation between $E_\mathrm{cyc}$ and
$L_\mathrm{X}$. For very small luminosities, the deceleration is
dominated by the gas pressure. In this regime larger mass accretion
rates will compress the accretion mound and thus lead to a positive
correlation between the luminosity and the cyclotron line energy.
Finally, at intermediate luminosities one expects a regime where the
accretion column is fairly stable and the deceleration is dominated by
Coulomb interactions.

\begin{figure}
 \includegraphics[width=\columnwidth]{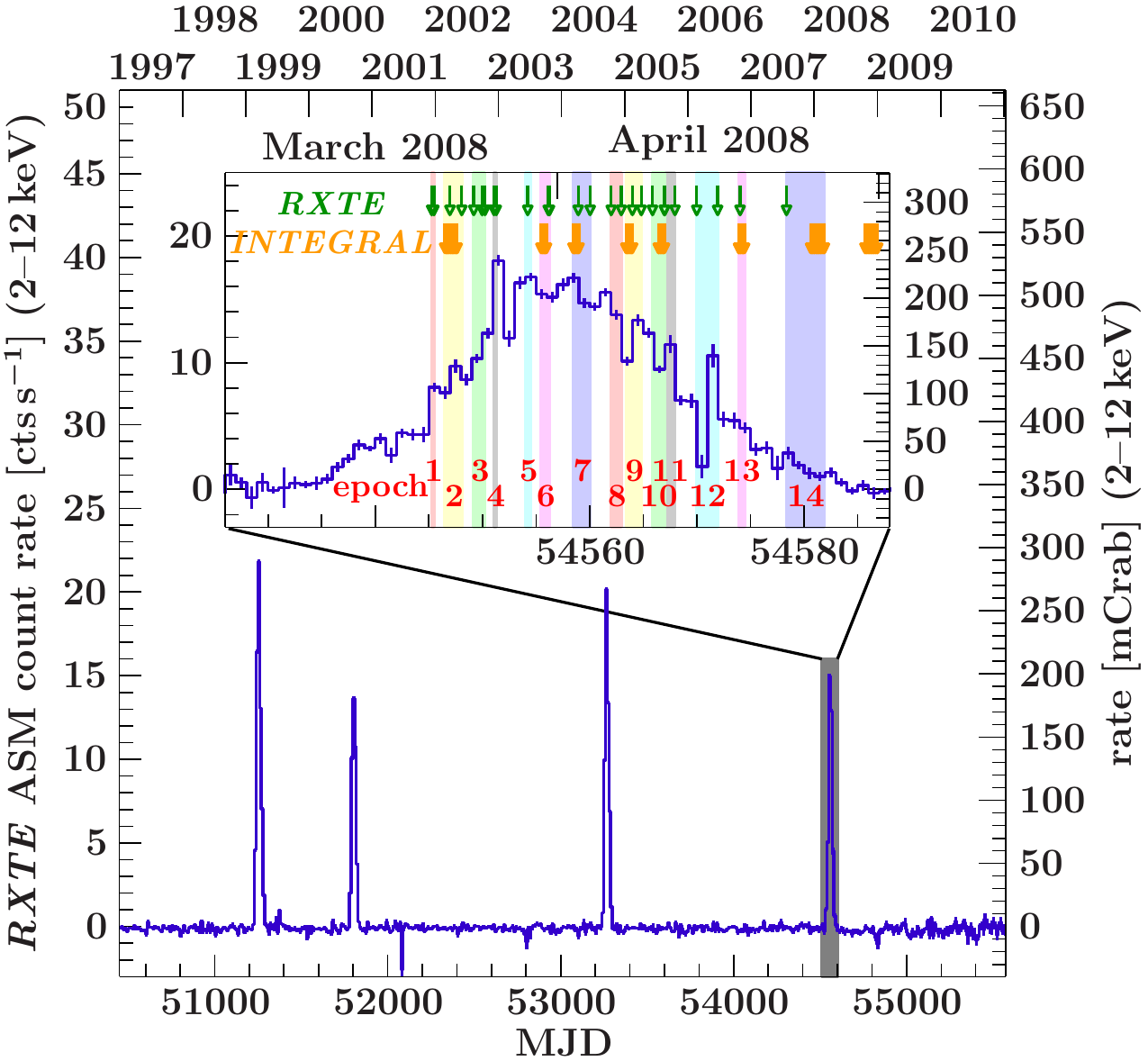}
 \caption{\rxte/ASM lightcurve of \object{4U~0115+634}
   (2--10\,keV). The inset shows a close-up view of this lightcurve,
   showing the 2008 outburst in detail. Arrows indicate the times of
   pointed \rxte\ and \textit{INTEGRAL} observations. Numbers indicate
   the observation epochs used in the data analysis (see also
   Table~\ref{obs}).}
 \label{fig:lc}
\end{figure}

\begin{table}
  \caption{Log of observations for the 2008 outburst of
    \object{4U~0115+634}. The first part of the table contains the
    \textit{RXTE} observations, the second part the ISGRI data.}\label{obs} 
\resizebox{\columnwidth}{!}{%
\begin{tabular}{llllr}
\hline\hline
ID\tablefootmark{a} & start date & MJD &
$t_\mathrm{exp}$\tablefootmark{b} [s]&
e\tablefootmark{c}\\
\hline
01-01-00 & 2008 Mar 20 & 54545.27--54545.43 & 8336/2413  &  1\\
01-01-01 & 2008 Mar 20 & 54545.47--54545.49 & 1648/382  &  1\\
01-02-04 & 2008 Mar 21 & 54546.97--54547.20 & 12448/3938 &  2\\
01-02-00 & 2008 Mar 23 & 54548.07--54548.25 & 9904/3082  &  2\\
01-02-01 & 2008 Mar 24 & 54549.19--54549.30 & 6336/1948  &  3\\
01-02-02 & 2008 Mar 24 & 54549.97--54550.14 & 8832/3136  &  3\\
01-02-03 & 2008 Mar 25 & 54550.24--54550.28 & 2384/934  &  3\\
01-02-05 & 2008 Mar 26 & 54551.08--54551.17 & 4096/1679  &  4\\
01-02-06 & 2008 Mar 26 & 54551.28--54551.31 & 752/625   &  4\\
01-03-02 & 2008 Mar 29 & 54554.23--54554.38 & 576/2112   &  5\\
01-03-01 & 2008 Mar 31 & 54556.14--54556.16 & 768/473   &  6\\
01-03-00 & 2008 Mar 31 & 54556.25--54556.37 & 768/1558   &  6\\
01-03-03 & 2008 Apr 02 & 54558.93--54559.10 & 5456/3104  &  7\\
01-04-00 & 2008 Apr 04 & 54560.04--54560.22 & 4240/2998 &  7\\
01-04-01 & 2008 Apr 06 & 54562.01--54562.18 & 5888/2908  &  8\\
01-04-02 & 2008 Apr 06 & 54562.95--54563.09 & 5552/2317  &  8\\
01-04-03 & 2008 Apr 07 & 54563.99--54564.07 & 3856/1548  &  9\\
01-04-04 & 2008 Apr 08 & 54564.82--54564.98 & 7808/2853  &  9\\
01-04-05 & 2008 Apr 09 & 54565.86--54566.10 & 10896/4101 & 10\\
01-04-06 & 2008 Apr 10 & 54566.98--54567.00 & 1648/699  & 10\\
01-05-00 & 2008 Apr 11 & 54567.00--54567.01 & 1056/381  & 11\\
01-05-01 & 2008 Apr 11 & 54567.96--54568.13 & 9632/3176  & 11\\
16-01-00 & 2008 Apr 13 & 54569.98--54570.16 & 10192/3043 & 12\\
16-01-01 & 2008 Apr 15 & 54571.94--54572.12 & 10368/3162 & 12\\
16-02-00 & 2008 Apr 18 & 54574.04--54574.15 & 7184/1924  & 13\\
16-02-01 & 2008 Apr 22 & 54578.36--54578.51 & 6480/1679  & 14\\
\hline
664 & 2008 Mar 21 & 54546.48--54547.68 & 55981&  2 \\ 
667 & 2008 Mar 30 & 54555.46--54556.04 & 28279&  6 \\ 
668 & 2008 Apr 02 & 54558.45--54559.05 & 27655&  7 \\ 
669 & 2008 Apr 07 & 54563.38--54564.02 & 28495&  9 \\ 
670 & 2008 Apr 10 & 54566.41--54567.01 & 30990&  10\\ 
673 & 2008 Apr 17 & 54573.90--54574.50 & 31360&  13\\ 
675 & 2008 Apr 24 & 54580.62--54581.97 & 70902&  14\\ 
677 & 2008 Apr 29 & 54585.69--54586.91 & 63662& --\\ 
\hline
\end{tabular}}
\tablefoot{
  \tablefoottext{a}{For \textit{RXTE}, the first column contains the
    number of the Obs-ID after ``93032-''. For ISGRI, it contains the
    revolution number.}
  \tablefootmark{b}{For \textit{RXTE}, the two numbers correspond to
    the PCA and HEXTE exposures, respectively.}
  \tablefoottext{c}{Epoch for data grouping, see text for details.}}
\end{table}

\object{4U~0115+634}\ \citep{Giacconi1972a} is one of the X-ray
pulsars for which CRSFs have been studied in great detail \citep[see,
e.g.,][]{Wheaton1979a,White1983a,
  Nagase1991a,Heindl1999a,Santangelo1999a,Mihara2004a,
  Nakajima2006a,Tsygankov2007a,Ferrigno2009a,Ferrigno2011a}. The
source shows 3.61\,s pulsation and has an orbital period of
$\sim$24.3\,d \citep{Cominsky1978a,Rappaport1978a}. The optical
counterpart is the O9e star V635~Cas \citep{Johns1978a,Unger1998a}.

In previous outbursts, CRSFs have been detected up to the fifth
harmonic \citep{Heindl1999a}. This is the largest number of detected
CRSFs in an accreting X-ray pulsar. In later work, a negative
correlation between $L_\mathrm{X}$ and $E_\mathrm{cyc}$ was found
\citep[e.g.,][]{Nakajima2006a,Tsygankov2007a,Mueller2010b}, meaning that
the source should be located in the supercritical luminosity regime.
Other authors even found a discontinuous behavior of the fundamental
line energy and explained this by a sudden change of the accretion
conditions \citep{Li2012a}.

In this work we revisit the behavior of the CRSFs in
\object{4U~0115+634}. Focusing on data taken during a giant outburst
in 2008 spring (Fig.~\ref{fig:lc}). We show that the previously
claimed relation between the souce flux $F_\mathrm{X}$ and
$E_\mathrm{cyc}$ strongly depends on the choice of the underlying
continuum model. In Sect.~\ref{sec:obs} we summarize the
observations. In Sect.~\ref{sec:specana} we describe the different
continuum models and the flux and time dependence of the continuum and
cyclotron line parameters. We argue that the cyclotron line is only
modelled correctly when using an exponentially cutoff power law
continuum with a strong emission feature around 10\,keV. Furthermore,
alternative continua models, which have been used in some earlier
investigations, result in incorrect cyclotron line parameters which
can cause the apparent relation between source flux $F_\text{X}$ and
fundamental cyclotron line energy $E_0$. We summarize this paper in
Sect.~\ref{sec:summary}.

\section{Observations and data reduction}\label{sec:obs}

The outburst of \object{4U~0115+634}\ started 2008 March 12, after a
phase of almost four years of quiescence, when \textit{Swift}/BAT
detected a significant increase of the X-ray flux \citep{Krimm2008a}.
The outburst lasted about 40\,days and exceeded a flux of
$\sim$200\,mCrab in the 2--12\,keV \textit{RXTE}/ASM band
(Fig.~\ref{fig:lc}). The whole outburst was monitored with pointed
observations by the Rossi X-ray Timing Explorer
\citep[\textit{RXTE},][]{Bradt1993a} and several pointed observations
by the INTErnational Gamma-Ray Astrophysics Laboratory
\citep[\textit{INTEGRAL},][]{Winkler2003a}. Here, we use data from
\textit{RXTE}'s Proportional Counter Array \citep[PCA,][]{Jahoda2006a}
and the High Energy X-Ray Timing Experiment
\citep[HEXTE,][]{Rothschild1998a}, as well as the \textit{INTEGRAL}
Soft Gamma-Ray Imager \citep[ISGRI,][]{Lebrun2003a}. Data were reduced
with our standard analysis pipelines, based on HEASOFT (v.~6.11) and
\textit{INTEGRAL} OSA v.~9.0. Data modeling was performed with the
newest version of the \textit{Interactive Spectral Interpretation
  System} \citep[ISIS,][]{Houck2000a}.

The PCA consisted of five proportional counter units (PCUs), which
were sensitive betweeen 2 and 60\,keV. Since PCU~2 is known to have
been the best calibrated one \citep{Jahoda2006a}, we only extracted
spectra from the top layer of this PCU in the \texttt{standard2f}
mode. The hard X-ray spectrum from 15--250\,keV was monitored with
HEXTE. HEXTE consisted of two independent clusters, A and B, both
arrays of four phoswich scintillation detectors. The ``rocking
mechanism'' of these clusters allowed a near realtime measurement of
the background. Since this mechanism failed for cluster~A before our
observations, we only used data from cluster B. The other instrument
for the hard X-rays was the pixelized CdTe dector ISGRI behind a Coded
Mask on board the \textit{INTEGRAL} satellite. It covers the energy
range between $\sim$18\,keV and 1\,MeV. We extracted all available
ISGRI spectra, for which \object{4U~0115+634}\ was less than
$10^\circ$ off-axis. Table~\ref{obs} contains an observation log (see
also Fig.~\ref{fig:lc}).

For the spectral analysis we used data between 3 and 50\,keV for PCA,
and between 20 and 100\,keV for HEXTE and for ISGRI. A systematic
error has been added in quadrature to the PCA and ISGRI data, using
the canonical values of 0.5\% and 2.0\% \citep[][and IBIS Analysis
User
Manual\footnote{http://www.isdc.unige.ch/integral/analysis\#Documentation}]
{Jahoda2006a}, respectively. To improve the signal to noise ratio of
the spectra, we averaged the data over 14 data blocks in time. These
data blocks are defined in the fifth column of Table~\ref{obs} and
shown in Fig.~\ref{fig:lc}. We omitted one ISGRI dataset from our
analysis (revolution number 677), because no simultaneous
\textit{RXTE} data are available at that time and the source was
almost in the off state during that observation.

\section{Spectral analysis}\label{sec:specana}

\subsection{Pulsar Continuum Models}\label{sec:continua}
A large problem when analyzing X-ray spectra from X-ray binary pulsars
is to find a good description of the broadband continuum. In most
cases, semi-physical models, which are some variant of a power law
with a high energy exponential cutoff \citep[][and references
therein]{Kreykenbohm2002a}, are used to describe the spectra. These
models can be justified by considering the Comptonization of soft
photons in an accretion column \citep[][and references
therein]{Becker2007a}. The most basic exponentially cutoff power law
model, called \texttt{\texttt{CutoffPL}} in ISIS and XSPEC
\citep{Arnaud1996a}, has the form
\begin{equation}\label{eq:cutoffpl}
  \mathrm{\texttt{CutoffPL}}(E)\propto E^{-\Gamma}\exp\left(-E/\Efold\right),
\end{equation}
with the photon index, $\Gamma$, and the folding energy, \Efold\
\citep[see, e.g.,][]{Schoenherr2007a}. However, this simple model is
often, but not always, insufficient to describe the continua of 
observed X-ray pulsars, as the curvature in the continuum is often
seen to start at different energies. To obtain a good spectral fit,
therefore, models with more free fit parameters and more complex
shapes are sometimes required. The most simple of these models is the
power law with high energy cutoff
\begin{equation}\label{eq:plcut}
\texttt{PLCUT}(E) \propto 
\begin{cases}
E^{-\Gamma}  & \mbox{for $E\le E_\mathrm{break}$ } \\
E^{-\Gamma} \exp\left((E_\mathrm{break}-E)/\Efold\right) & \mbox{for
    $E>E_\mathrm{break}$}
\end{cases}
\end{equation}
where for typical X-ray pulsars $E_\mathrm{break}\sim 10$\,keV and
which has an unphysical ``kink'' at $E_\mathrm{break}$, which can
result in line-like residuals around this energy
\citep{Kretschmar1997a,Kreykenbohm1999a}. For this reason
\citet{Klochkov2008a} and \citet{Ferrigno2009a} smooth the continuum
around this break, using a model of the form
\begin{equation}\label{eq:bkn}
\texttt{PLINT}(E) \propto 
\begin{cases}
E^{-\Gamma} & \mbox{for $E\le E_\mathrm{break}-\Delta E$} \\
AE^3 + BE^2 +CE +D & \mbox{for $|E-E_\mathrm{break}|<\Delta E$} \\
E^{-\Gamma}\exp\left(-E/\Efold\right) & \mbox{for $E\ge
  E_\mathrm{break}+\Delta E$}
\end{cases}
\end{equation}
where the coefficients $A$, $B$, $C$, and $D$ are chosen such that the
continuum and its first derivative are continuous at
$E=E_\mathrm{break}\pm\Delta E$ and where $\Delta E\ne 0$ but
sufficiently small.

Other common choices to avoid the break are the so called Negative and
Positive power-law Exponential \citep[\texttt{NPEX}, see, e.g.,][and
references therein]{Makishima1999a}, a continuum model with five free
parameters of the form
\begin{equation}\label{eq:npex}
\mathrm{\texttt{NPEX}}(E) \propto \left(E^{-\Gamma_1}+\alpha
  E^{+\Gamma_2}\right) \exp\left(-E/\Efold\right),
\end{equation}
with photon indices $\Gamma_{1,2}\ge 0$\footnote{Often one fixes
  $\Gamma_2=2$, see, e.g., \citet{Nakajima2006a}.}, and the ``Fermi
Dirac Cutoff'' \citep[\texttt{FDCUT}, see][]{Tanaka1986a}
\begin{equation}\label{eq:fdcut}
\texttt{FDCUT}(E) \propto E^{-\Gamma} \,
\frac{1}{1+\exp\left((E-E_\mathrm{break})/\Efold\right)}. 
\end{equation}
Common to all of these empirical continua is that despite the smoother
transition around $E_\mathrm{break}$ they often need to be modified
further by a broad excess or depression around 10\,keV. This so-called
``10\,keV feature'' has been observed in many X-ray pulsar spectra,
either in absorption or in emission, although its origin is still not
clear \citep{Coburn2001a}. This feature is usually modeled as an
additive Gaussian line, with the centroid energy $E_\text{G}$, the
width $\sigma_\text{G}$, and the flux $A_\text{G}$.

In addition to those empirical models, based on the theoretical
description of bulk and thermal Comptonization in magnetized accretion
colums of \citet[][and references therein]{Becker2007a},
\cite{Ferrigno2009a} have developed a continuum model depending on
physical parameters like the culumn radius, the accretion rate, the
electron temperature, the $B$-field strength. A hybrid model which
combines the broadband continuum with CRSFs is currently in
preparation \citep{SchwarmPrep}. These physics-based continuum models
are still in active development and, because of the large number of
free parameters, only suited for very high signal to noise data. For
this reason most observational work uses one of the continua described
by Eq.~\ref{eq:cutoffpl}--\ref{eq:fdcut}.

For sources with cyclotron lines, the continuum model is modified with
a description of the cyclotron line. Since physically self consistent
line models such as those described by \citet{Schoenherr2007a} are
still not available for all sources, empirical line models are
generally used. The two most common CRSF models are those using a
Gaussian or a pseudo-Lorentzian optical depth profile \citep[see
discussion by][]{Nakajima2010a}. Here, we model the CRSFs with a
pseudo-Lorentzian optical depth profile \citep{Mihara1990a}, i.e., we
multiply the continuum model with
\begin{equation}\label{eq:CYCLABS}
\texttt{CYCLABS}(E)=\exp\left(-\frac{\tau(W E/E_\mathrm{cyc})^2}{(E-E_\mathrm{cyc})^2+W^2}\right),
\end{equation}
with the centroid energy, $E_\mathrm{cyc}$, the width of the feature,
$W$, and the optical depth of the line, $\tau$. This approach allows
the comparison of the line parameters with results from previous
papers, in which also the \texttt{CYCLABS} model has been used
\citep[e.g.,][]{Nakajima2006a,Tsygankov2007a,Li2012a}. In the
remainder of the paper, we label the cyclotron line parameters with
the number of respective harmonic, where 0 corresponds to the
fundamental line. As common for CRSFs, $W$ and $\tau$ are strongly
correlated \citep{Coburn2001a}. We therefore set a lower limit of
0.5\,keV for the width, which is comparable to the typical PCA
resolution.

In order to describe the data of \object{4U~0115+634}, in addition to
several of the continuum models discussed above, we take interstellar
absorption into account, modelling it with an updated version of
TBabs\footnote{see
  http://pulsar.sternwarte.uni-erlangen.de/wilms/research/tbabs/},
using abundances by \citet{Wilms2000a} and cross sections by
\citet{Verner1995a}. We furthermore account for an intrinsic Fe
K$\alpha$ flourescence line, which we describe with an additive narrow
Gaussian emission line with frozen centroid energy at 6.4\,keV, a
width of $10^{-4}$\,keV, and a flux $A_\text{Fe}$ and equivalent width
$W_\text{Fe}$. Assuming a narrow line is justified because in systems
like 4U~0115+63 Fe K$\alpha$ is expected to have a width in the order
of $\lesssim$0.5\,keV \citep[see, e.g.,][]{Torrejon2010a}, which
cannot be resolved by PCA.

In order to take into account that the flux normalization of the
different instruments used is not perfectly known we introduce cross
calibration constants $c_\mathrm{HEXTE}$ and $c_\mathrm{ISGRI}$ as
free fit parameters. These constants also account for flux differences
of the source between the observations which were not fully
simultaneous. The PCA background was estimated using the standard
models provided by GSFC. These estimates often show a flux deviation
from the proper background by a few percent. This deviation was
accounted for by multiplying the background flux with another fit
parameter, $c_\mathrm{b}$ \citep[see also][]{Rothschild2011a}.

\begin{figure}
 \includegraphics[width=\columnwidth]{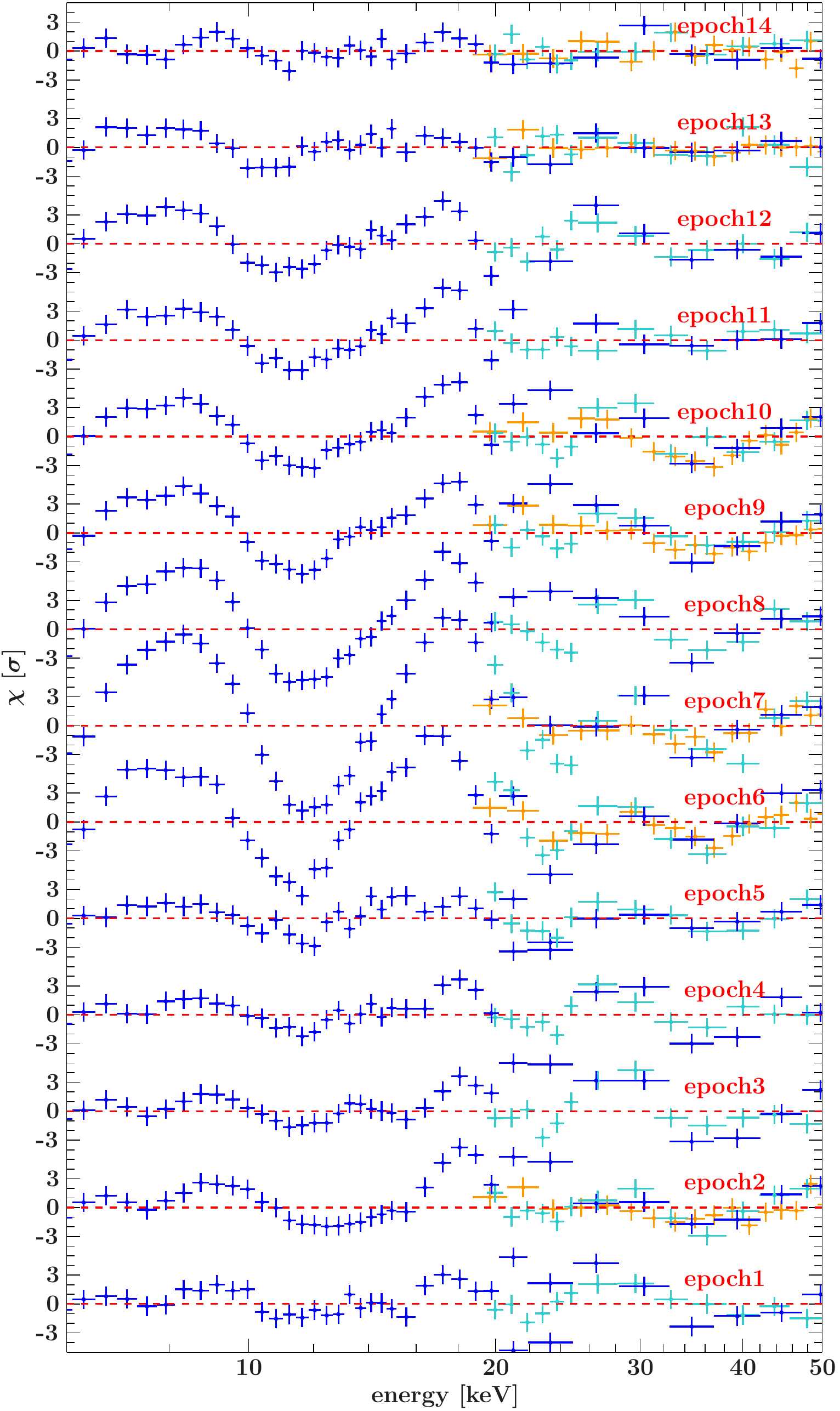}
 \caption{Deviations between the data and the best fit with
   \texttt{CutoffPL} and 10\,keV feature for the PCA (circles), HEXTE
   (crosses), and ISGRI (filled diamonds) in units of standard
   deviations $\sigma$. For these fits, no cyclotron lines have been
   taken into account. The numbers on the right side indicate the
   respective epoch.}
 \label{fig:residuals}
\end{figure}

\begin{figure}
\includegraphics[width=\columnwidth]{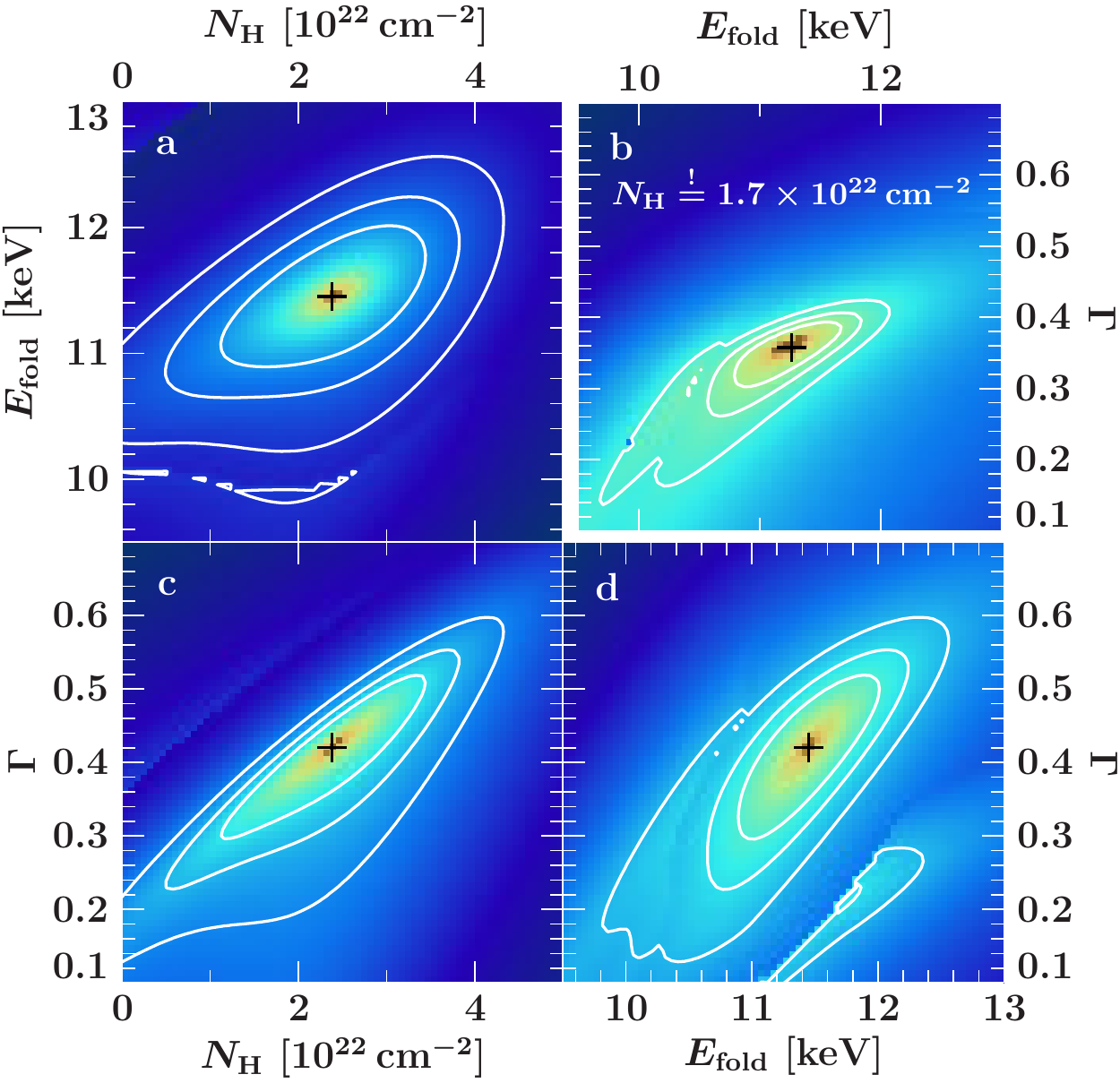}
\caption{\textbf{a}, \textbf{c}, and \textbf{d}: Confidence contours
  for the folding energy, the photon index and the hydrogen column
  density for epoch~6. b: Confidence contours between the \Efold\ and
  $\Gamma$ with $N_\text{H}$ frozen to
  \mbox{$1.7\times10^{22}\,\text{cm}^{-2}$}. Contour lines correspond
  to the 68.3\%, 90\%, and 99\% level. Color indicates $\Delta\chi^2$
  with respect to the best fit value, with the color scale running
  from orange (low $\Delta\chi^2$) to dark blue (large
  $\Delta\chi^2$). }\label{fig:contconf}
\end{figure}

\begin{figure}
\includegraphics[width=\columnwidth]{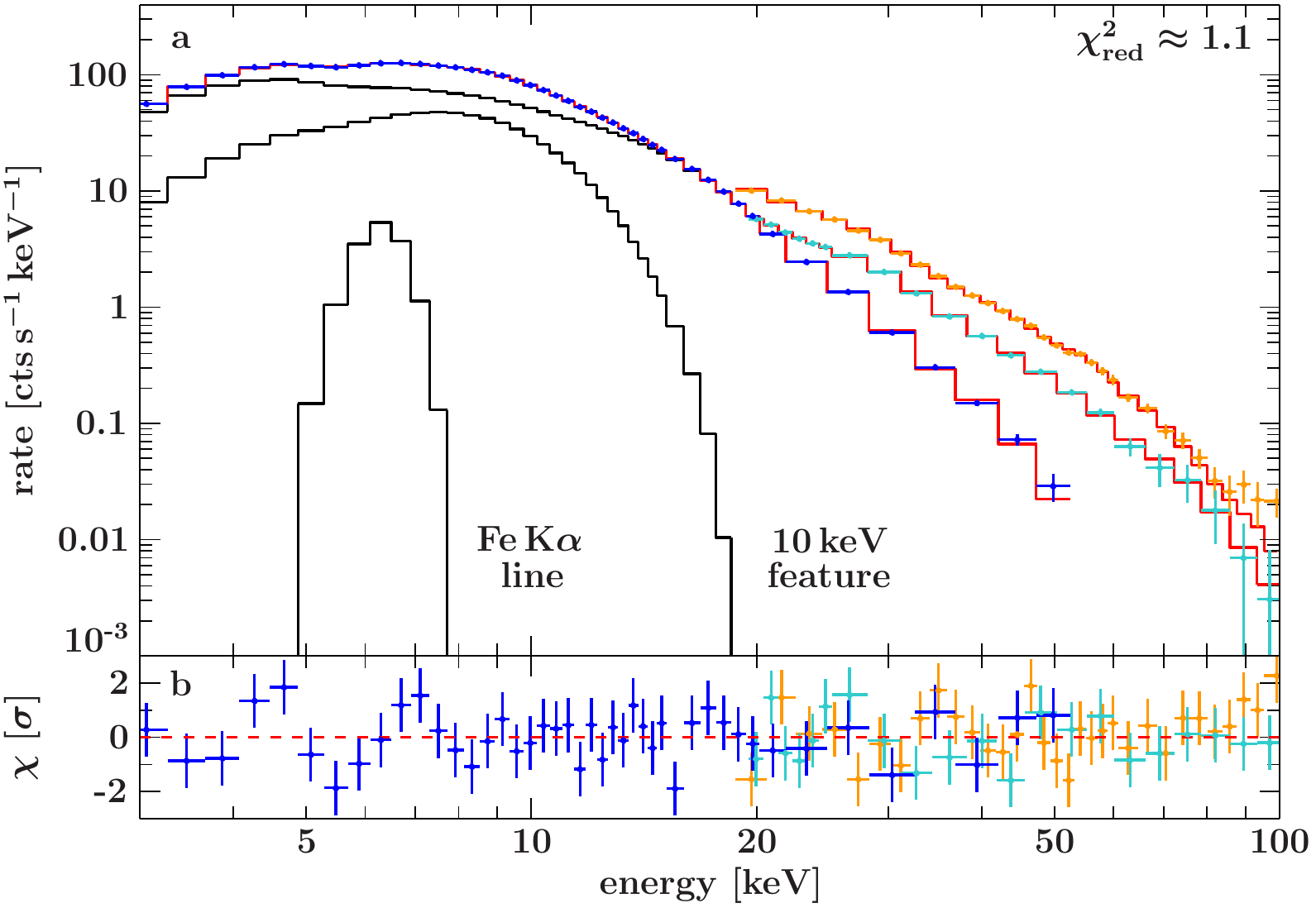}
\caption{\textbf{a}: Folded spectrum from epoch 6 together with best
  fit (red) and all model components, i.e., \texttt{CutoffPL}, 10\,keV
  feature, Fe K$\alpha$ flourescence line, and five cyclotron
  features. \textbf{b}: Residuals of the best fit, shown as the
  difference between the data and the model normalized to the
  uncertainty of each data point. Dark blue data points correspond to
  PCA, cyan points to HEXTE, and orange points to ISGRI. }
\label{fig:bestfit}
\end{figure}

\subsection{Fitting Strategy}

Since the \texttt{CutoffPL} is the easiest continuum model for X-ray
pulsars, we used this model for our fits. Furthermore, the 10\,keV
feature in emission is required to get a good description of the data.
For each epoch, the fits have been simultaneously performed for all
available intstruments. Figure~\ref{fig:residuals} shows the residuals
of the best fit using this continuum without taking any CRSFs into
account. These residuals clearly indicate the need of the cyclotron
lines in this model at \mbox{$\sim$11}, 22, and sometimes above
30\,keV. As the quality of individual data sets is strongly variable
over the outburst, in an initial fit run a variable number of CRSFs
were added to individual fits until the reduced $\chi^2$,
$\chi^2_\mathrm{red}\sim 1$. To avoid continuum modeling with the
absorption features, we fixed the widths of the CRSFs of the second
and higher harmonics to typical values, i.e., 4\,keV
\citep{Ferrigno2009a}.

In order to check whether the number of different cyclotron features
in the model affects the fit results of the continuum parameters and
the fundamental and first harmonic CRSF, in a second run all spectral
fits included the fundamental line and its first and second harmonic
only. For epochs, where only the first harmonic was necessary to
obtain a good fit in our initial run, we added the second harmonic
with fixed parameters during the second run, holding its parameters at
the mean values of the results from the other epochs.

Since initial fits with the \texttt{CutoffPL} continuum showed large
uncertainties for the hydrogen column density, $N_\mathrm{H}$, without
a significant variability between different epochs, we held
$N_\mathrm{H}$ fixed at the mean value $N_\mathrm{H}=1.7\times
10^{22}\,\mathrm{cm}^{-2}$ to avoid unphysical correlations between
these fit parameters (Fig.~\ref{fig:contconf}).

\begin{figure}
%jw having this figure wider than this breaks latex
 \includegraphics[width=0.9\columnwidth]{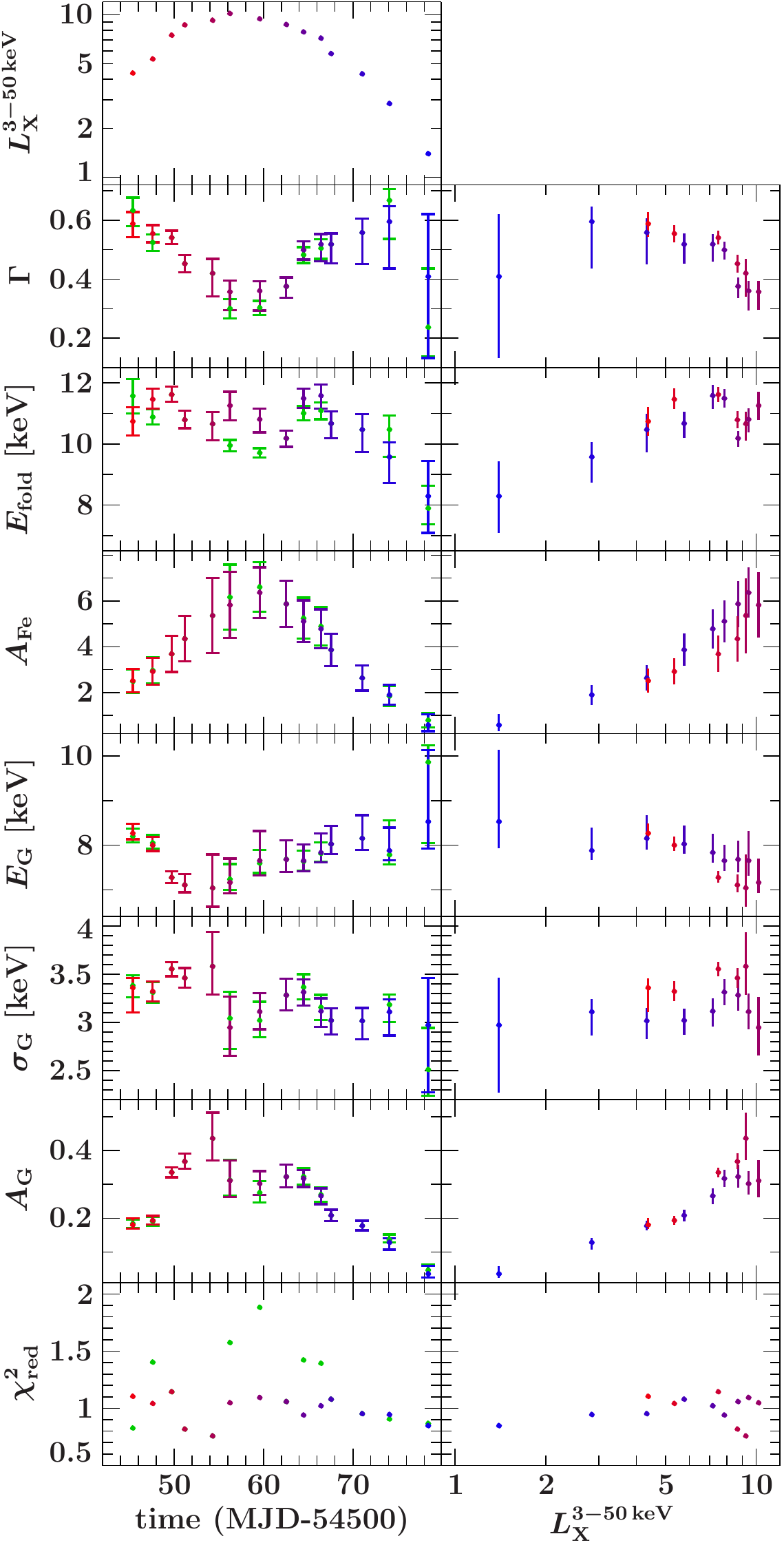}
 \caption{\textbf{Left} Temporal evolution and \textbf{right}
   luminosity dependence of the continuum parameters (see text for
   definitions) and 10\,keV feature for the fits with the
   \texttt{CutoffPL} continuum. The colour gradient from reddish to
   blueish data points indicates the temporal evolution of the
   outburst. The data points show results from the initial fit run,
   where a variable number of cyclotron lines was used. Green data
   points in the left column correspond to the second fit run with a
   fixed number of three cyclotron lines (only in the left column),
   which was been performed as a consistency check (see text for
   details). Where no green points are visible, both fits gave the
   same results. $L_\text{X}$ is in units of
   $10^{37}\,\text{erg}\,\text{s}^{-1}$. $A_\text{G}$ and
   $A_\text{Fe}$ are in units of
   $\text{photons}\,\text{s}^{-1}\,\text{cm}^{-2}$, while
   $A_\text{Fe}$ is in multiples of $10^{-3}$. }\label{fig:contpar}
\end{figure}

The two fit runs show that almost all fit parameters are equal within
their respective uncertainties for both runs. Only \Efold\ and the
line parameters of the second harmonic are slightly different in
epoch~6 and~7. These differences can be explained by the fact that for
these two cases the second run has to account for the equivalent of
the four harmonics required in the first run. The differences affect
mainly the higher energies and are compensated by a change in the
folding energy and the closest cyclotron line, i.e., the second
harmonic. The remainder of this paper is based on the results from the
first fit run.

An example for the best fit using the \texttt{CutoffPL} continuum with
10\,keV feature and five cyclotron features (fit run a, epoch~6) is
shown together with all model components in
Fig.~\ref{fig:bestfit}. This spectrum was recorded by all three
instruments during the maximum of the outburst, i.e., epoch 6. All fit
results from runs using the \texttt{CutoffPL} continuum are summarized
in Table~\ref{tab:cont}.

\begin{table*}
  \caption{Continuum parameters of the time resolved spectral
    analysis. Fits were performed with the \texttt{CutoffPL} model and a
    varying number of CRSFs.}\label{tab:cont} 
\renewcommand{\arraystretch}{1.3}
\begin{tabular}{lllllllllllllll}
\hline\hline
ep\tablefootmark{a} & $L_\mathrm{X}$\tablefootmark{b}  & $\Gamma$ & \Efold &
$A_\mathrm{Fe}$\tablefootmark{c}&$W_\mathrm{Fe}$ & $E_\mathrm{G}$ &
$A_\mathrm{G}$\tablefootmark{c} &$\sigma_\mathrm{G}$ & $c_\mathrm{b}$&  $c_\mathrm{HEXTE}$
& $c_\mathrm{ISGRI}$ & \chisq/dof\\
 &  &  &[keV] & $[10^{-3}]$&[eV] & [keV]&  &  [keV] &  &  &  \\
\hline
$\phantom{0}$1    & $\phantom{0}4.38^{+0.02}_{-0.02}$   & $0.59^{+0.04}_{-0.05}$ & $10.7^{+0.5}_{-0.5}$ & $2.5^{+0.6}_{-0.6}$& $55^{+12}_{-12}$ & $8.3^{+0.3}_{-0.2}$ & $0.18^{+0.02}_{-0.02}$ & $3.4^{+0.1}_{-0.3}$ & $1.00^{+0.02}_{-0.02}$ & $0.73^{+0.02}_{-0.02}$ & --&1.13/50\\
$\phantom{0}$2    & $\phantom{0}5.35^{+0.02}_{-0.02}$   & $0.56^{+0.03}_{-0.03}$ & $11.5^{+0.4}_{-0.4}$ & $2.9^{+0.7}_{-0.7}$& $53^{+10}_{-10}$ & $8.0^{+0.2}_{-0.2}$ & $0.19^{+0.02}_{-0.02}$ & $3.3^{+0.1}_{-0.1}$ & $0.96^{+0.01}_{-0.01}$ & $0.74^{+0.01}_{-0.01}$ & $0.87^{+0.02}_{-0.02}$& 1.12/77\\
$\phantom{0}$3    & $\phantom{0}7.48^{+0.02}_{-0.02}$   & $0.54^{+0.03}_{-0.03}$ & $11.6^{+0.3}_{-0.3}$ & $3.7^{+0.9}_{-0.9}$&$46^{+10}_{-10}$  & $7.3^{+0.2}_{-0.2}$ & $0.34^{+0.02}_{-0.01}$ & $3.6^{+0.1}_{-0.1}$ & $0.94^{+0.02}_{-0.02}$ & $0.75^{+0.01}_{-0.01}$ & --&1.32/48\\
$\phantom{0}$4    & $\phantom{0}8.66^{+0.03}_{-0.03}$   & $0.45^{+0.03}_{-0.03}$ & $10.8^{+0.3}_{-0.3}$ & $4.4^{+1.0}_{-1.0}$& $46^{+10}_{-10}$ & $7.1^{+0.3}_{-0.2}$ & $0.37^{+0.03}_{-0.03}$ & $3.5^{+0.1}_{-0.1}$ & $0.94^{+0.02}_{-0.02}$ & $0.78^{+0.01}_{-0.01}$ & --&0.88/48\\
$\phantom{0}$5    & $\phantom{0}9.24^{+0.05}_{-0.06}$   & $0.42^{+0.05}_{-0.08}$ & $10.7^{+0.4}_{-0.6}$ & $5.4^{+1.7}_{-1.7}$& $52^{+17}_{-17}$ & $7.0^{+0.8}_{-0.5}$ & $0.44^{+0.08}_{-0.07}$ & $3.6^{+0.4}_{-0.3}$ & $0.92^{+0.05}_{-0.05}$ & $0.94^{+0.02}_{-0.02}$ & --&0.79/48\\
$\phantom{0}$6    & $10.18^{+0.04}_{-0.04}$ & $0.36^{+0.04}_{-0.07}$ & $11.3^{+0.5}_{-0.6}$ & $5.8^{+1.6}_{-1.5}$&$51^{+13}_{-13}$  & $7.2^{+0.6}_{-0.3}$ & $0.31^{+0.07}_{-0.05}$ & $3.0^{+0.4}_{-0.3}$ & $0.93^{+0.04}_{-0.04}$ & $0.88^{+0.02}_{-0.01}$ & $1.04^{+0.02}_{-0.02}$&  1.09/75\\
$\phantom{0}$7    & $\phantom{0}9.43^{+0.03}_{-0.03}$   & $0.36^{+0.04}_{-0.07}$ & $10.8^{+0.4}_{-0.5}$ & $6.4^{+1.2}_{-1.2}$& $61^{+10}_{-10}$ & $7.7^{+0.7}_{-0.4}$ & $0.30^{+0.04}_{-0.04}$ & $3.1^{+0.2}_{-0.2}$ & $0.94^{+0.02}_{-0.02}$ & $0.84^{+0.01}_{-0.01}$ & $1.05^{+0.02}_{-0.02}$& 1.18/75\\
$\phantom{0}$8    & $\phantom{0}8.70^{+0.03}_{-0.03}$ & $0.38^{+0.04}_{-0.04}$ & $10.2^{+0.3}_{-0.3}$ & $5.9^{+1.0}_{-1.0}$&$61^{+10}_{-10}$  & $7.7^{+0.4}_{-0.3}$ & $0.32^{+0.04}_{-0.04}$ & $3.3^{+0.2}_{-0.2}$ & $0.93^{+0.02}_{-0.02}$ & $0.78^{+0.01}_{-0.01}$ & --&1.27/48\\
$\phantom{0}$9    & $\phantom{0}7.84^{+0.03}_{-0.02}$   & $0.50^{+0.03}_{-0.04}$ & $11.5^{+0.4}_{-0.3}$ & $5.1^{+1.0}_{-1.0}$& $60^{+10}_{-10}$ & $7.7^{+0.4}_{-0.3}$ & $0.32^{+0.03}_{-0.03}$ & $3.3^{+0.2}_{-0.2}$ & $0.93^{+0.02}_{-0.02}$ & $0.75^{+0.01}_{-0.01}$ & $0.96^{+0.02}_{-0.02}$& 0.99/77\\
10  & $\phantom{0}7.17^{+0.03}_{-0.02}$   & $0.52^{+0.04}_{-0.07}$ & $11.6^{+0.4}_{-0.5}$ & $4.8^{+0.9}_{-1.0}$& $61^{+12}_{-12}$ & $7.8^{+0.5}_{-0.3}$ & $0.27^{+0.03}_{-0.03}$ & $3.1^{+0.2}_{-0.2}$ & $0.95^{+0.02}_{-0.02}$ & $0.75^{+0.01}_{-0.01}$ & $0.87^{+0.02}_{-0.02}$&  1.11/77\\
11  & $\phantom{0}5.76^{+0.02}_{-0.03}$   & $0.52^{+0.04}_{-0.07}$ & $10.7^{+0.4}_{-0.5}$ & $3.9^{+0.8}_{-0.8}$& $61^{+12}_{-12}$ & $8.0^{+0.4}_{-0.3}$ & $0.21^{+0.02}_{-0.02}$ & $3.0^{+0.2}_{-0.2}$ & $0.97^{+0.02}_{-0.02}$ & $0.76^{+0.01}_{-0.01}$ & --&1.17/48\\
12  & $\phantom{0}4.33^{+0.02}_{-0.02}$   & $0.56^{+0.06}_{-0.10}$ & $10.5^{+0.6}_{-0.7}$ & $2.6^{+0.6}_{-0.6}$&$53^{+12}_{-12}$  & $8.2^{+0.5}_{-0.3}$ & $0.18^{+0.02}_{-0.02}$ & $3.0^{+0.2}_{-0.2}$ & $0.96^{+0.01}_{-0.01}$ & $0.73^{+0.01}_{-0.01}$ & --&1.10/48\\
13  & $\phantom{0}2.84^{+0.02}_{-0.02}$   & $0.59^{+0.06}_{-0.16}$ & $\phantom{0}9.6^{+0.5}_{-0.9}$ & $1.9^{+0.5}_{-0.5}$&$55^{+13}_{-14}$  & $7.9^{+0.6}_{-0.3}$ & $0.13^{+0.02}_{-0.03}$ & $3.1^{+0.2}_{-0.3}$ & $0.95^{+0.02}_{-0.02}$  & $0.72^{+0.02}_{-0.02}$ & $0.86^{+0.03}_{-0.03}$& 0.96/81\\
14  & $\phantom{0}1.40^{+0.02}_{-0.02}$   & $0.41^{+0.22}_{-0.28}$ & $\phantom{0}8.3^{+1.2}_{-1.2}$ & $0.6^{+0.5}_{-0.3}$&$35^{+27}_{-17}$  & $8.5^{+0.5}_{-0.7}$ & $0.04^{+0.03}_{-0.02}$ & $3.0^{+0.5}_{-0.7}$ & $0.98^{+0.02}_{-0.02}$  & $0.69^{+0.04}_{-0.04}$ & $0.44^{+0.03}_{-0.03}$& 0.86/81\\
\hline\hline
ep\tablefootmark{a} & $E_0$ & $W_0$& $\tau_0$&  $E_1$ &
$W_1$ & $\tau_1$&  $E_2$ & $\tau_2$&  $E_3$ &
$\tau_3$& $E_4$ & $\tau_4$\\
& [keV] & [keV]& & [keV] & [keV]& &[keV] & & [keV] & &[keV] & \\
\hline
$\phantom{0}$1    & $11.0^{+0.4}_{-0.8}$ & $0.9^{+3.9}_{-0.4}$ & $0.04^{+0.02}_{-0.02}$ & $21.6^{+0.3}_{-0.3}$ & $0.9^{+1.0}_{-0.4}$ & $0.24^{+0.18}_{-0.07}$ & --            & --            & -- & -- & -- & --\\
$\phantom{0}$2    & $11.2^{+0.5}_{-0.6}$ & $0.8^{+2.4}_{-0.4}$ & $0.03^{+0.02}_{-0.02}$ & $22.1^{+0.3}_{-0.3}$ & $2.6^{+1.0}_{-0.9}$ & $0.24^{+0.03}_{-0.03}$ & $32.7^{+1.3}_{-1.3}$ & $0.25^{+0.07}_{-0.07}$ & $42.8^{+2.5}_{-2.2}$ & $0.26^{+0.09}_{-0.08}$ & -- & --\\
$\phantom{0}$3    & $11.1^{+0.5}_{-0.5}$ & $0.5^{+1.3}_{-0.0}$ & $0.03^{+0.02}_{-0.02}$ & $22.0^{+0.2}_{-0.2}$ & $0.5^{+0.8}_{-0.0}$ & $0.26^{+0.05}_{-0.12}$ & $37.5^{+1.0}_{-1.0}$ & $0.20^{+0.05}_{-0.05}$ & -- & -- & -- & --\\
$\phantom{0}$4    & $11.0^{+0.5}_{-0.7}$ & $0.8^{+2.4}_{-0.3}$ & $0.03^{+0.02}_{-0.02}$ & $21.8^{+0.3}_{-0.3}$ & $1.1^{+0.9}_{-0.6}$ & $0.19^{+0.16}_{-0.05}$ & $36.4^{+1.0}_{-1.2}$ & $0.22^{+0.07}_{-0.07}$ & -- & -- & -- & --\\
$\phantom{0}$5    & $11.0^{+0.6}_{-0.7}$ & $2.5^{+1.5}_{-1.5}$ & $0.12^{+0.15}_{-0.07}$ & $22.1^{+0.6}_{-0.6}$ & $1.4^{+2.3}_{-0.9}$ & $0.18^{+0.24}_{-0.07}$ & $36.2^{+1.8}_{-2.0}$ & $0.18^{+0.09}_{-0.08}$ & -- & -- & -- & --\\
$\phantom{0}$6    & $10.7^{+0.3}_{-0.4}$ & $1.8^{+1.0}_{-1.2}$ & $0.11^{+0.14}_{-0.07}$ & $21.8^{+0.4}_{-0.4}$ & $4.8^{+0.9}_{-0.9}$ & $0.46^{+0.05}_{-0.05}$ & $35.2^{+0.7}_{-1.0}$ & $0.48^{+0.07}_{-0.08}$ & $46.7^{+1.2}_{-2.0}$ & $0.44^{+0.10}_{-0.13}$ & $59.7^{+2.0}_{-3.2}$ & $0.52^{+0.24}_{-0.25}$\\
$\phantom{0}$7    & $10.7^{+0.2}_{-0.3}$ & $2.3^{+0.9}_{-0.9}$ & $0.18^{+0.14}_{-0.08}$ & $22.1^{+0.2}_{-0.2}$ & $3.8^{+0.6}_{-0.5}$ & $0.45^{+0.05}_{-0.03}$ & $33.3^{+0.9}_{-0.8}$ & $0.35^{+0.06}_{-0.07}$ & $41.6^{+1.4}_{-1.6}$ & $0.34^{+0.08}_{-0.08}$ & $52.8^{+1.8}_{-2.0}$ & $0.42^{+0.13}_{-0.14}$\\
$\phantom{0}$8    & $10.6^{+0.3}_{-0.3}$ & $2.3^{+0.8}_{-0.7}$ & $0.16^{+0.08}_{-0.05}$ & $21.8^{+0.3}_{-0.3}$ & $2.9^{+0.8}_{-0.7}$ & $0.30^{+0.03}_{-0.04}$ & $34.9^{+1.0}_{-0.8}$ & $0.25^{+0.06}_{-0.05}$ & -- & -- & -- & --\\
$\phantom{0}$9    & $10.5^{+0.3}_{-0.3}$ & $2.1^{+0.9}_{-0.7}$ & $0.12^{+0.07}_{-0.04}$ & $21.6^{+0.3}_{-0.4}$ & $2.7^{+1.2}_{-0.9}$ & $0.22^{+0.04}_{-0.03}$ & $33.2^{+1.2}_{-1.0}$ & $0.25^{+0.06}_{-0.06}$ & $43.7^{+2.0}_{-1.7}$ & $0.28^{+0.08}_{-0.07}$ & -- & --\\
10  & $10.3^{+0.4}_{-0.5}$ & $2.0^{+1.2}_{-0.9}$ & $0.09^{+0.10}_{-0.04}$ & $21.4^{+0.4}_{-0.5}$ & $3.9^{+1.4}_{-1.0}$ & $0.25^{+0.06}_{-0.04}$ & $34.0^{+1.3}_{-1.0}$ & $0.31^{+0.07}_{-0.07}$ & $44.0^{+2.3}_{-1.9}$ & $0.28^{+0.09}_{-0.08}$ & -- & --\\
11  & $10.3^{+0.4}_{-0.4}$ & $2.1^{+1.0}_{-1.0}$ & $0.10^{+0.10}_{-0.05}$ & $21.3^{+0.4}_{-0.4}$ & $2.9^{+1.6}_{-1.0}$ & $0.21^{+0.05}_{-0.03}$ & $33.0^{+2.4}_{-1.5}$ & $0.22^{+0.08}_{-0.07}$ & -- & -- & -- & --\\
12  & $10.1^{+0.3}_{-0.3}$ & $2.3^{+1.0}_{-0.9}$ & $0.14^{+0.13}_{-0.06}$ & $20.8^{+0.3}_{-0.4}$ & $2.3^{+1.0}_{-0.8}$ & $0.21^{+0.05}_{-0.04}$ & $34.6^{+1.8}_{-1.9}$ & $0.20^{+0.10}_{-0.08}$ & -- & -- & -- & --\\
13  & $10.2^{+0.4}_{-0.5}$ & $1.7^{+1.7}_{-1.0}$ & $0.08^{+0.13}_{-0.03}$ & $20.7^{+1.4}_{-2.2}$ & $2.6^{+5.0}_{-2.1}$ & $0.08^{+0.13}_{-0.05}$ & --            & --               & -- & -- & -- & --\\
14  & $10.7^{+0.5}_{-0.8}$ & $0.5^{+4.0}_{-0.0}$ & $0.06^{+0.04}_{-0.03}$ & $20.3^{+1.6}_{-1.4}$ & $5.9^{+2.8}_{-3.2}$ & $0.38^{+0.19}_{-0.19}$ & --            & --               & -- & -- & -- & --\\
\hline
\end{tabular}
\tablefoot{
  Uncertainties and upper limits are at the 90\% confidence level for
  one interesting parameter.\\
  \tablefoottext{a}{Epoch for data grouping (see text for details).}
  \tablefoottext{b}{Absorbed luminosity in units of $10^{37}\,\text{erg\,s}^{-1}$. $L_\text{X}$ covers the energy band 3--50\,keV and was
    calculated using a distance of 7\,kpc \citep{Negueruela2001a}.}
  \tablefoottext{c}{In units of $\text{photons}\,\text{s}^{-1}\,\text{cm}^{-2}$.}}
\end{table*}

\section{Spectral Evolution: No Anti-Correlation of the Cyclotron Line
  Energy with Flux}

\begin{figure}
 \includegraphics[width=\columnwidth]{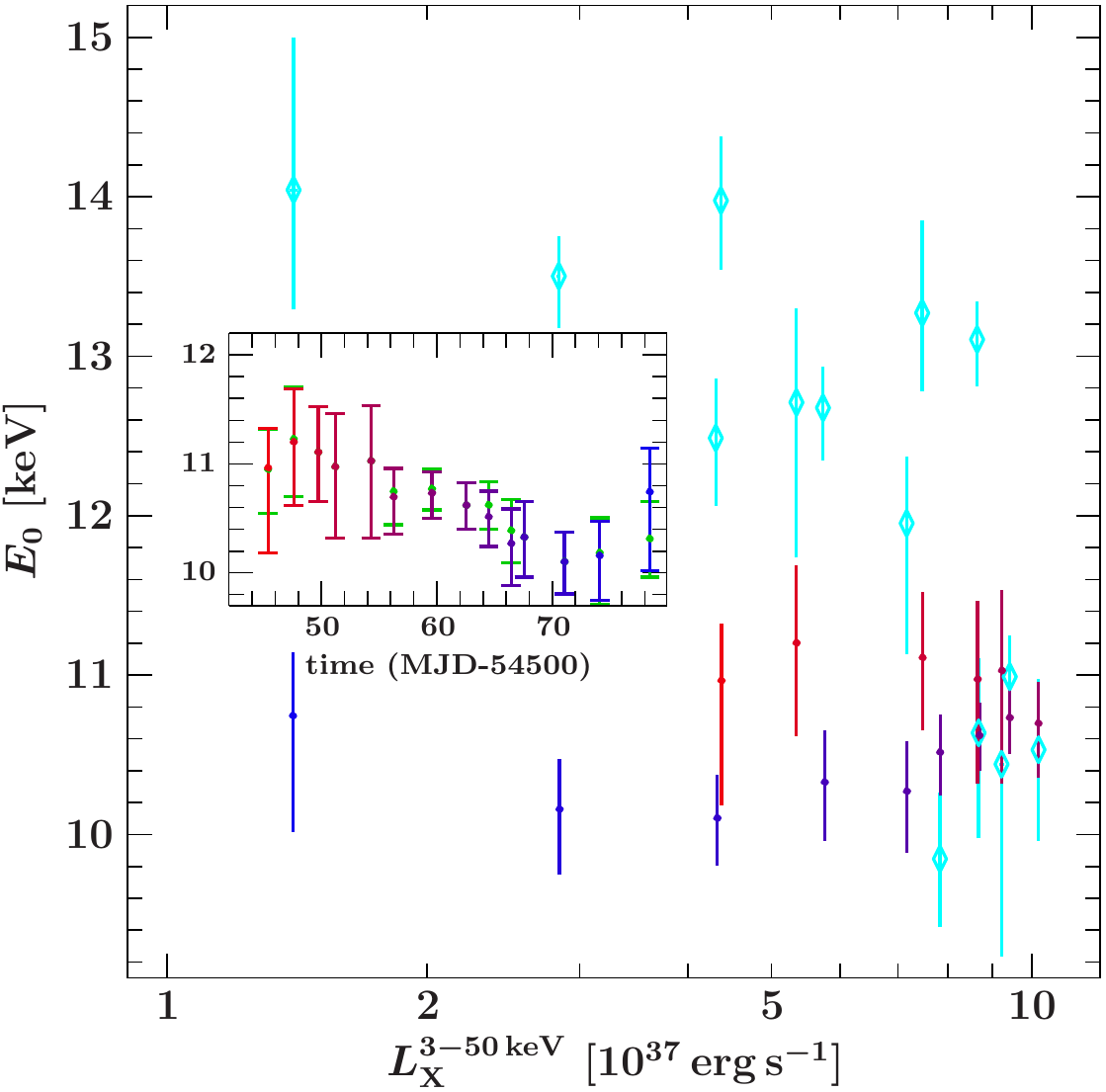}
 \caption{Luminosity and time dependence (inset) of the energy of the
   fundamental CRSF using the \texttt{CutoffPL} continuum model. The
   color coding is the same as in Fig.~\ref{fig:contpar}. Cyan data
   points (diamonds) show the results when using the \texttt{NPEX}
   continuum.}\label{fig:E0}
\end{figure}

\begin{figure}
 \includegraphics[width=\columnwidth]{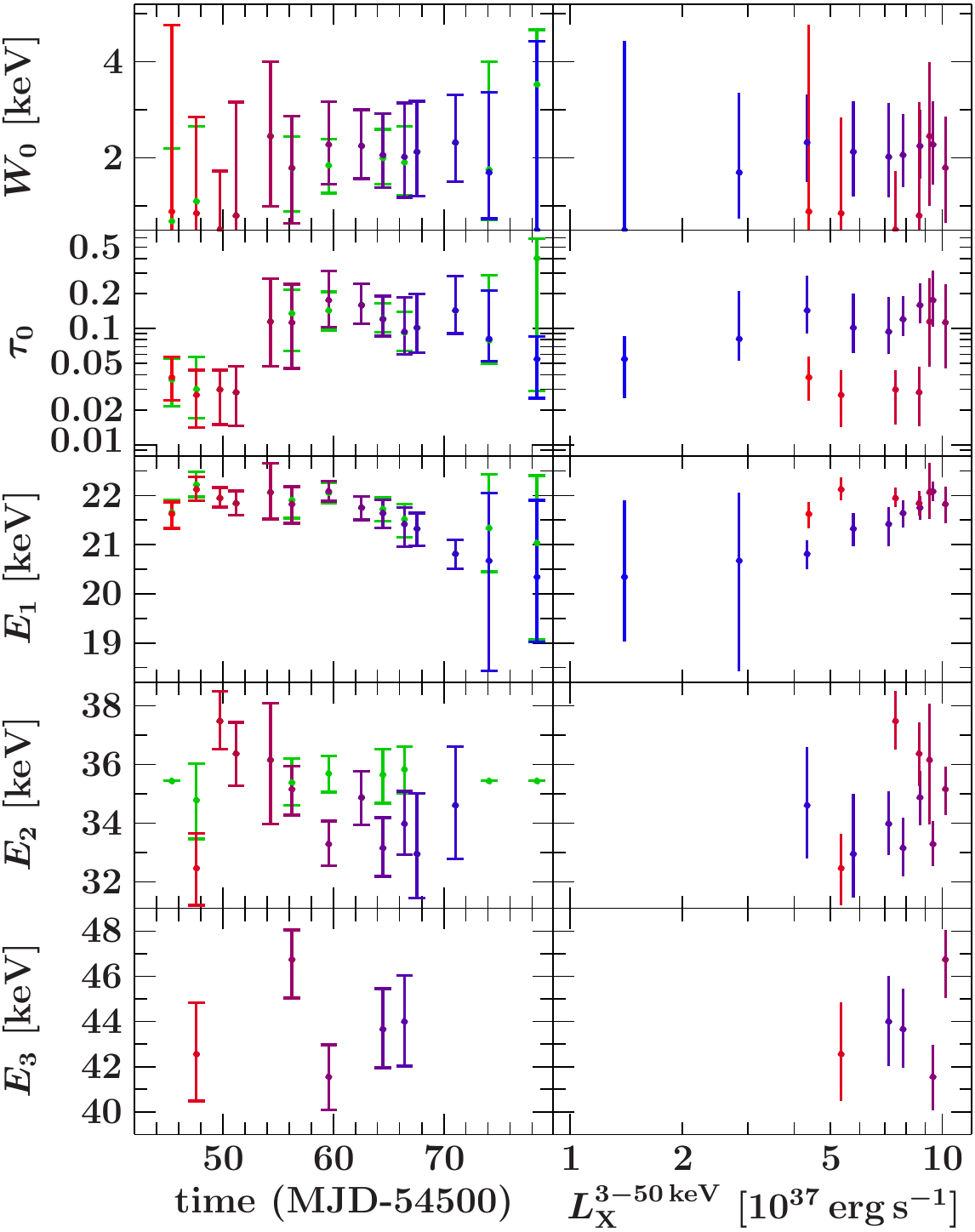}
 \caption{Time (left) and luminosity (right) dependence of the
   parameters of the CRSF. The colour coding is the same as in
   Fig.~\ref{fig:contpar}.}\label{fig:WT0}
\end{figure}

Figures~\ref{fig:contpar}--\ref{fig:WT0} show the evolution of the
continuum and cyclotron line parameters as well as the parameters for
the Fe K$\alpha$ line for the the first fit run with the
\texttt{CutoffPL} continuum and 10\,keV feature, and fixed
$N_\mathrm{H}$ (see also Tab. \ref{tab:cont}). As already noted in
previous analyses for \object{4U~0115+634}, the continuum varies over
the outburst \citep[e.g.,][]{Li2012a}. During the brightest phases of
the outburst the spectrum is the hardest one. As shown in
Fig.~\ref{fig:contconf}, the variation of $\Gamma$ is significantly
larger than the very slight correlation seen between $\Gamma$ and
$E_\mathrm{fold}$. As expected, the source's intrinsic iron K$\alpha$
line is significantly present during all epochs. As also observed in
other X-ray transient pulsars \citep{Inoue1985a} and expected for a
line originating from fluorescence, the line flux is positively
correlated to the X-ray flux from the source. The equivalent width
stays constant within the uncertainties over the outburst. The 10\,keV
feature varies in both, the centroid energy $E_\text{G}$ and its flux
$A_\text{G}$. Its energy seems to be slightly anti-correlated with the
source flux, while the relative flux in the feature, i.e., the ratio
between the flux in the the 10 keV feature and the total flux, remains
constant. Its width, $\sigma_\mathrm{G}$, remains relatively constant
between 3.0\,keV and 3.5\,keV.

Very different from virtually all previous analyses, however, the
cyclotron line is seen to be only slightly varying over the outburst.
The colored data points in Fig.~\ref{fig:E0} and \ref{fig:WT0} show
the centroid energy, $E_0$, the width, $W_0$, and the optical depth,
$\tau_0$, of the fundamental cyclotron line against time and flux for
the \texttt{CutoffPL} continuum, modified with the 10\,keV feature. In
these fits, the width $W_0$ remains constant throughout the outburst,
and there is only a slight ($\sim$$3\sigma$) indication that the
optical depth $\tau_0$ was somewhat shallower during the initial
phases of the outburst, before $\sim$MJD~54552, although it cannot be
excluded that some of this effect is due to a correlation with the
``10\,keV feature'', which during this time of the outburst peaks
close to the cyclotron line at around 8\,keV. Most importantly,
however, as shown in Fig.~\ref{fig:E0} the energy of the fundamental
cyclotron line is only very slightly variable: during early phases of
the outburst the line is around 11\,keV, moving towards 10\,keV as the
outburst progressed. This result is contrary to all earlier studies of
outbursts of \object{4U~0115+634}, where strong changes in the line
energy were seen
\citep[e.g.,][]{Li2012a,Tsygankov2007a,Nakajima2006a,Mihara2004a}.

Where does this difference in line behavior come from? As discussed in
Sect.~\ref{sec:continua} a large variety of continuum models can be
used to describe the broad band spectra of accreting neutron stars.
Due to the early successes of fitting spectra of \object{4U~0115+634}
with the \texttt{NPEX} and \texttt{PLCUT} models, these continua have
been extensively used when modeling the outburst behavior of this
pulsar. Unfortunately, however, in most previous analyses -- including
some of our own -- no attempt was made at modeling the data with any
of the other available models.

As an example, modeling the 2008 outburst of \object{4U~0115+634} with
the \texttt{NPEX} model \citep[following][fixing $\Gamma_2$ to
2.0]{Nakajima2006a} results in fits that are of only slight worse
quality as those using the \texttt{CutoffPL} model modified by the
10\,keV feature, and we can recover the strong variation of the
cyclotron line found in earlier analyses (Fig.~\ref{fig:E0}, cyan data
points)\footnote{Modeling the spectra with the \texttt{PLCUT} model
  results in an unphysical break around the energy of the fundamental
  line, we therefore consider this continuum to be unsuitable for
  studying the cyclotron line behavior.}. Depending on the choice of
the continuum, we therefore find a fundamentally different behavior of
the fundamental cyclotron line, especially the centroid energy.

%jw - move to the suzaku paper, it's more appropriate there and
% the nice figure would be mainly ignored if we leave it in here
%
% \begin{figure}
%  \includegraphics[width=\columnwidth]{Fe}
%  \caption{Results for the flux and the equivalent width of the Fe
%    K$\alpha$ line. $A_\text{Fe}$ is in units of
%    $10^{-3}\,\text{photons}\,\text{s}^{-1}\,\text{cm}^{-2}$. The
%    colour coding is the same as in
%    Fig.~\ref{fig:contpar}.} \label{fig:Fe}
% \end{figure}

\begin{figure}
 \includegraphics[width=\columnwidth]{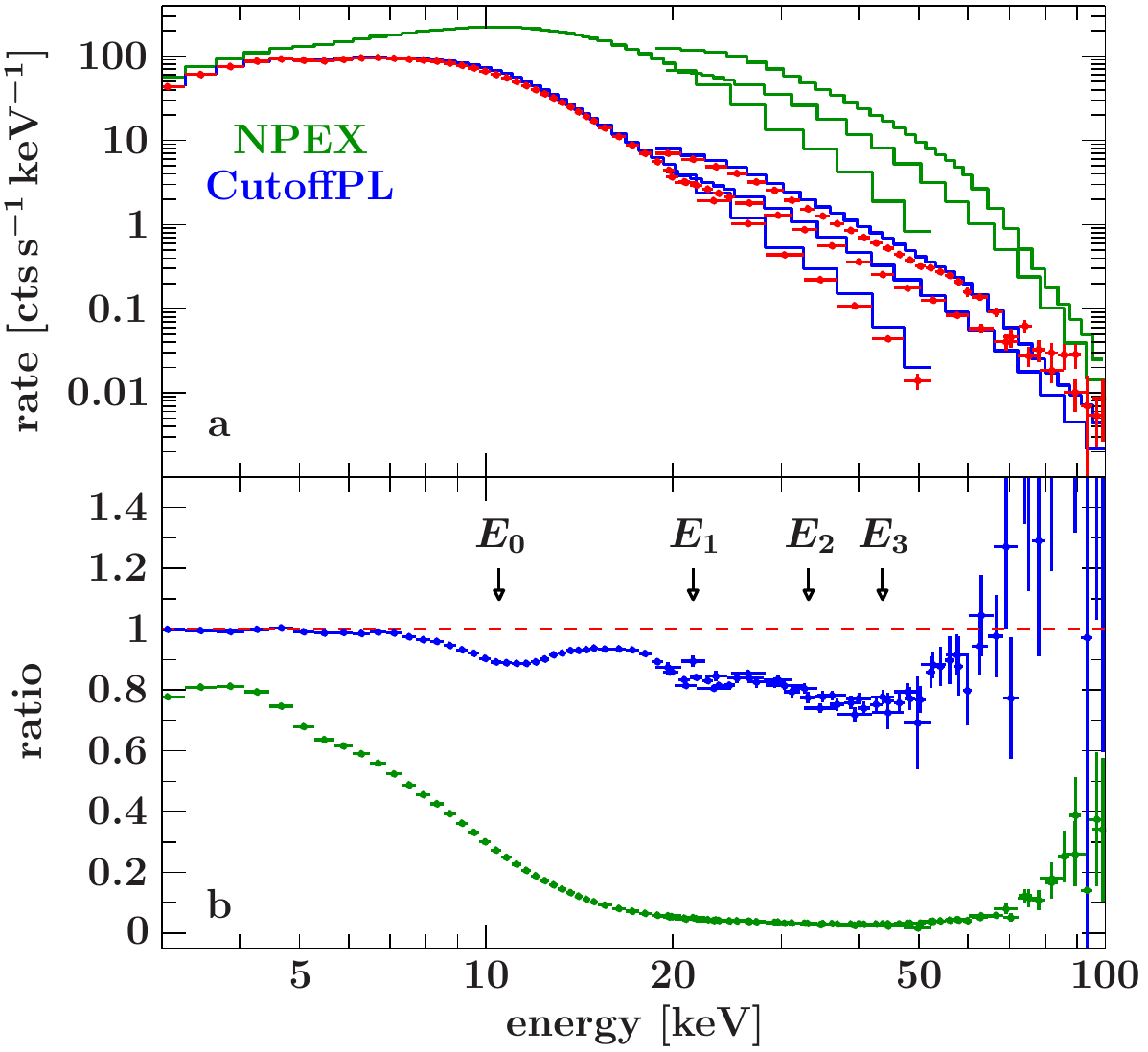}
 \caption{\textbf{a} PCA, HEXTE, and ISGRI spectra from epoch 9 (red)
   together with the best fit models using the \texttt{NPEX} (green) and
   \texttt{CutoffPL} (blue) continua after setting the optical depth
   of the cyclotron lines to zero. \textbf{b} Ratio between the data
   and the model.} \label{fig:tau0}
\end{figure}

Given that the fits with \texttt{NPEX} and \texttt{CutoffPL} give
similarly good results, what is the fundamental difference between
both modeling approaches? \texttt{NPEX} models generally result in
very broad cyclotron lines, with $W_0$ and $W_1$ often exceeding
5\,keV \citep[e.g.,][runs using the \texttt{PLCUT} model give similar
results, see, e.g., \citealt{Li2012a}]{Nakajima2006a}. In contrast,
using the \texttt{CutoffPL} continuum with the 10\,keV feature results
in widths of typically less than 3\,keV, much more consistent with the
narrow shapes of the residuals shown in Fig.~\ref{fig:residuals}. When
lines are as broad as in the \texttt{NPEX} fits, they can influence
the continuum fit. To illustrate how strongly these features actually
distort the continuum, Fig.~\ref{fig:tau0} shows the continuum shape
inferred when setting $\tau_0$ and $\tau_1$ to zero. The residuals for
the \texttt{CutoffPL} continuum with the 10\,keV feature (dark blue)
show sharp absorption dips at the cyclotron line energies of the
fundamental and first harmonic line energies. For the \texttt{NPEX}
model (green line), on the other hand, it is obvious that rather than
describing narrow cyclotron lines, the multiplicative broad line model
strongly influences the continuum. As discussed above, our spectral
fits show that during the brightest phases of the outburst a
significant hardening of the continuum emission is observed, together
with a change of the exponential cutoff. While this is also seen in
the variation of the continuum parameters, it is very likely that the
luminosity dependence of the CRSF in the \texttt{NPEX} fits is due to
the line partially modeling this behavior of the continuum, and not a
real physical effect.

\begin{figure}
 \includegraphics[width=\columnwidth]{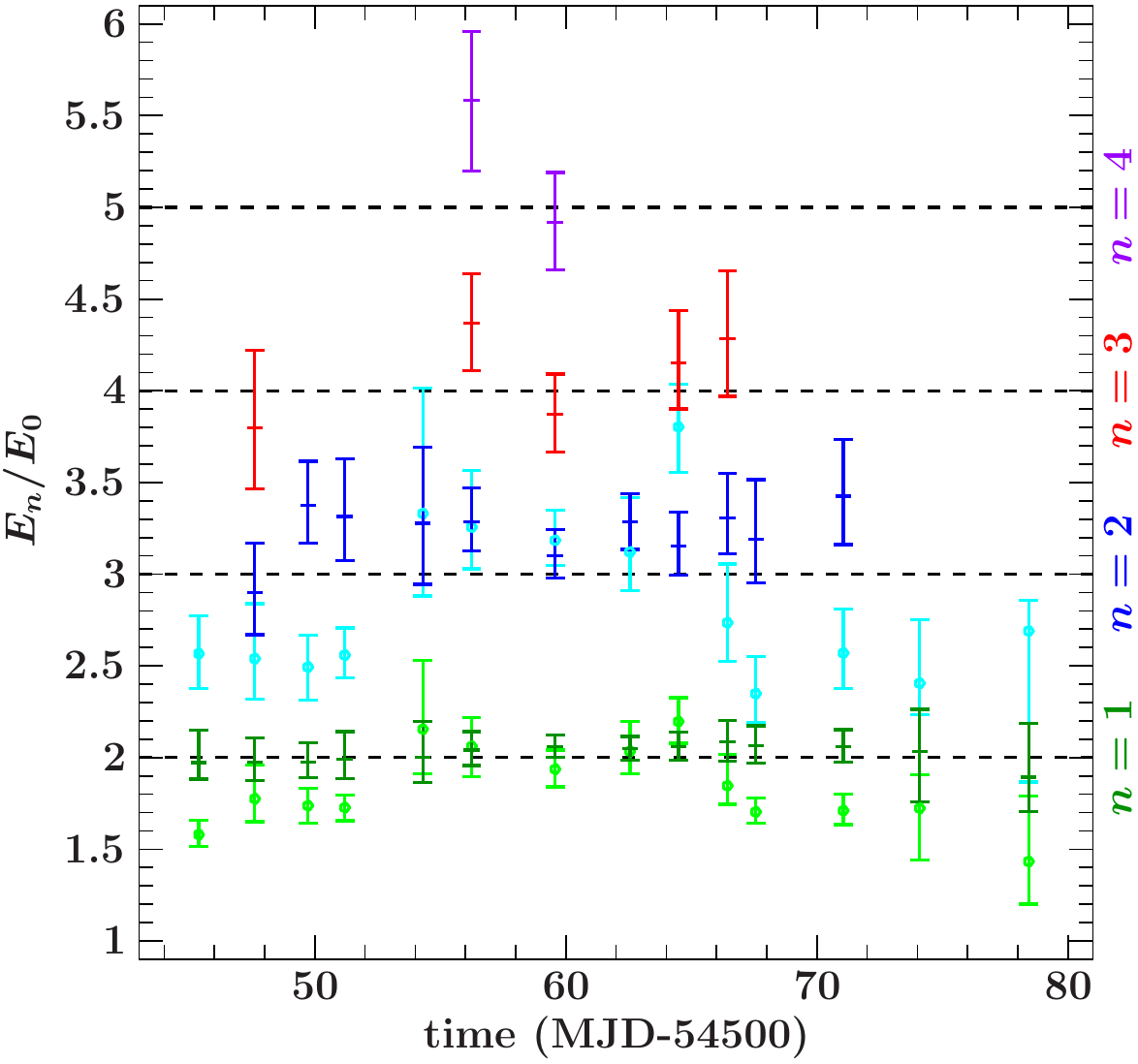}
 \caption{Ratios of the cyclotron line energies $E_n/E_0$ with respect
   to the centroid energy of the fundamental line against time. The
   dark green, dark blue, red and purple data points (bars) correspond
   to $E_1/E_0$, $E_2/E_0$, $E_3/E_0$, and $E_4/E_0$, respectively,
   using the \texttt{CutoffPL} continuum with a 10\,keV feature. The
   light green and cyan data points (filled circles) show the
   results for $E_1/E_0$ and $E_2/E_0$ using \texttt{NPEX}. The dashed
   lines indicate integer values for these ratios.}
 \label{fig:EvsE}
\end{figure}

Using the results from the \texttt{CutoffPL} plus 10\,keV feature fits
for the line energy also solves another problem found in earlier
spectral modeling. Here, using \texttt{NPEX} fits, the ratios between
the fundamental and higher cyclotron lines were often found to deviate
significantly from integers. While slightly non-integer ratios are
expected when taking relativistic quantum mechanics into account
\citep[see the discussion by][]{Pottschmidt2005a}, the large
deviations from integer multiples seen previously
\citep[e.g.,][]{Tsygankov2007a,Nakajima2006a,Heindl1999a,Santangelo1999a}
require rather complex model assumptions which introduce second order
effects such as strong vertical $B$-field gradients, crustal field
structures on small scales, or thermomagnetic effects for their
explanation \citep[see][and references therein for a
discussion]{Schoenherr2007a}. Figure~\ref{fig:EvsE} shows that we can
recover these non-integer ratios in our \texttt{NPEX} fits, but when
using the \texttt{CutoffPL} plus 10\,keV feature model, these ratios
are mostly in agreement with integer values, or at least only slightly
higher, as expected from the most simple models for cyclotron line
formation. We conclude that also from a physical point of view, the
\texttt{CutoffPL} continuum with the 10\,keV feature gives a much more
satisfactory description of the data.

\begin{figure}
 \includegraphics[width=\columnwidth]{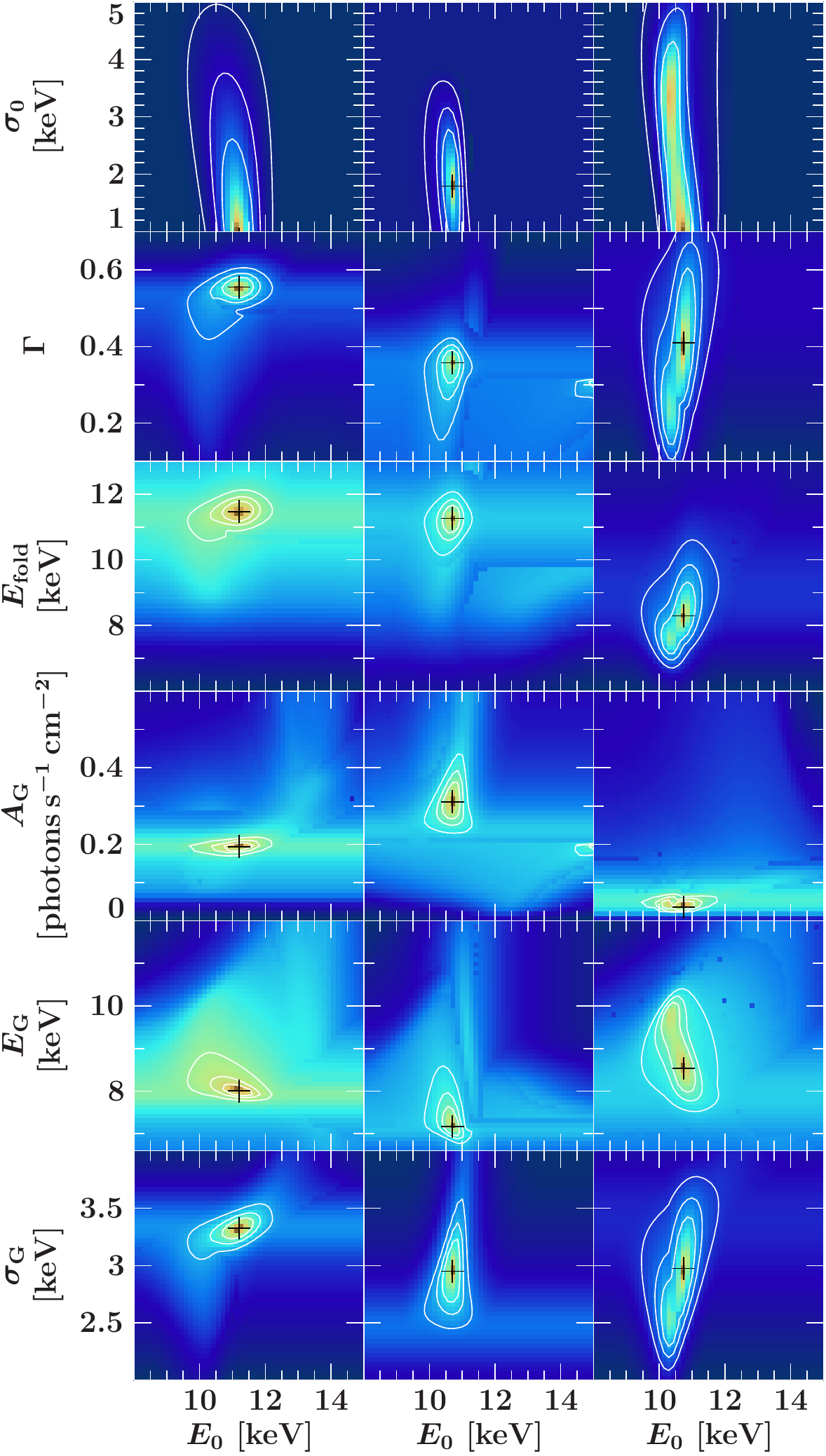}
 \caption{Confidence contours between the fundamental CRSF energy and
   other fit parameters. The columns show the results for epochs 2, 6,
   and 14, respectively. Contour lines correspond to the 68.3\%, 90\%,
   and 99\% confidence level. Color indicates $\Delta\chi^2$
  with respect to the best fit value, with the color scale running
  from orange (low $\Delta\chi^2$) to dark blue (large
  $\Delta\chi^2$). }
 \label{fig:cont5}
\end{figure}

Finally, we check the dependence of the line position on other free
fit parameters when using the \texttt{CutoffPL} continuum modified by
the 10\,keV feature. Since the origin of the 10\,keV feature remains
still an open questtion, possible relations between the CRSF centroid
energy and the parameters of the 10\,keV feature have to be
investigated, and other continuum parameters such as \Efold\ or
$\Gamma$ could influence the results for the cylotron line
parameters. Figure~\ref{fig:cont5} shows the behavior of
$\Delta\chi^2$ in the vicinity of the best fit values for three epochs
representative for the early phase, peak, and late phase of the 2008
outburst for selected combinations of important fit parameters. These
contour plots show that the line energy is not affected by cross
correlations with other parameters, i.e., possible variations in the
line energy of the fundamental cyclotron line are not due to
variations observed in the parameters of the 10\,keV feature or the
continuum parameters.

\begin{figure}
 \includegraphics[width=\columnwidth]{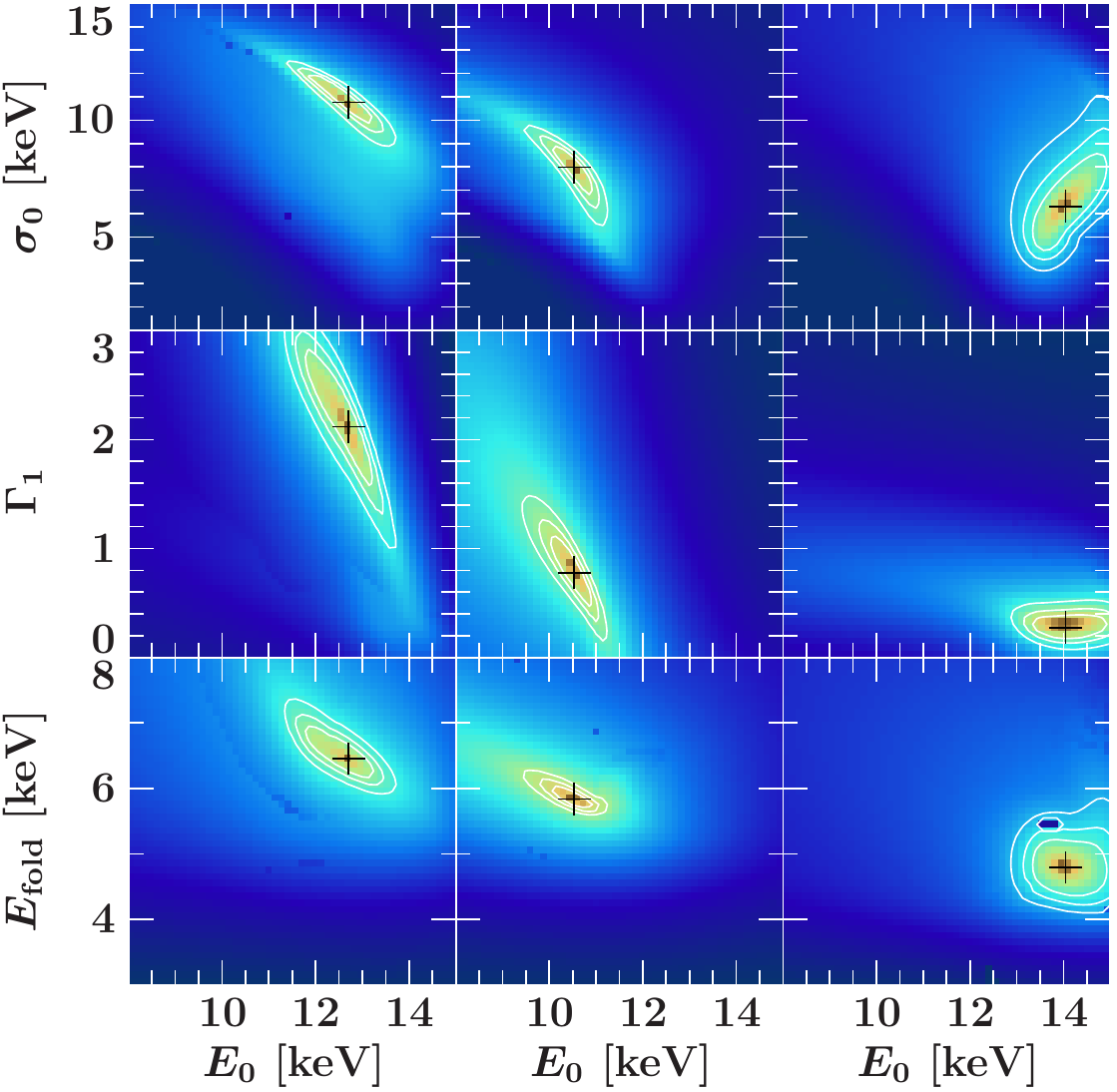}
 \caption{Confidence contours between the fundamental cyclotron line
   energy and continuum parameters when using the NPEX model. The
   displayed epochs, colours and lines are the same as in
   Fig.~\ref{fig:cont5}.}
 \label{fig:contNPEX}
\end{figure}

In contrast to the stable behavior of the line when using
\texttt{CutoffPL} with a 10\,keV feature, Fig.~\ref{fig:contNPEX}
illustrates the dependence of the CRSF centroid energy on the
continuum parameters of the NPEX model. In many cases, strong
correlations between $E_0$ and $\Gamma_1$ as well as between $E_0$ and
$E_\text{fold}$ are present. These correlations are a further
indicator that the Lorentzian component is used to model the continuum
rather than a cyclotron line when using the NPEX model.

\section{Summary and Conclusion}\label{sec:summary}

In this paper we have presented a study of the 2008 outburst of
\object{4U~0115+634} based on \textit{RXTE} and \textit{INTEGRAL} data
(Fig.~\ref{fig:lc}). We reproduced the results from previous work
\citep[e.g.,][and others]{Nakajima2006a,Tsygankov2007a,Li2012a}, that
the spectra can be successfully modelled by, e.g., the \texttt{NPEX}
model. We showed that for these fits the very broad absorption
features, which are thought to describe the CRSFs, rather model the
broadband continuum. Our result that the continuum parameters are
strongly variable over the outburst (Fig.~\ref{fig:contpar}), could be
responsible for the change in the cyclotron line energy in these fits.

We have shown that the spectral continuum can be also well described
with a model introduced by \citet{Klochkov2007a} for EXO~2030+375 and
\citet{Ferrigno2009a} for 4U~0115+634, i.e., the \texttt{CutoffPL},
modified by strong Gaussian emission feature around 10\,keV and
several (up to four) cyclotron lines (Table~\ref{tab:cont} and
Fig.~\ref{fig:bestfit}). Due to an unphysical break close to the
energy of the fundamental cyclotron line, the \texttt{PLCUT} model
does not give a physical description of the continuum. Using the
\texttt{CutoffPL} together with the 10\,keV feature, we have shown
that the modeling of lines reflects the theoretical expectation in
both the line shape and the ratios of the harmonics. In this model,
the luminosity dependency of the centroid energy is not present. As
shown also by \citet{Ferrigno2011a}, the line centroid energy is
remarkably stable throughout the brightest phase of the outburst. Here
we show that such stability is present, albeit at lower significance,
also during the early and late phases. In this model, the continuum
variations are explained by a slight anti-correlation of the centroid
energy of the Gaussian feature with luminosity and a significant
variation of $\Gamma$.

The strong systematic influence of the chosen continuum on the
cyclotron line behavior illustrates a problem that is potentially
inherent to all cyclotron line measurements. We emphasize, however,
that this does not mean that the general picture of the different
regimes of magnetized neutron star accretion and general luminosity
dependence of the cyclotron line as outlined by \citet[][and
references therein]{Becker2012a} is incorrect. In the sample of
sources discussed by \citet{Becker2012a}, \object{4U~0115+634} was an
outlier. The change of line energy inferred from the \texttt{NPEX} and
\texttt{PLCUT} models was almost a jump at a 3--50\,keV luminosity of
around $3\times 10^{37}\,\mathrm{erg}\,\mathrm{s}^{-1}$ and strong
hysteresis effects were present (Fig.~\ref{fig:E0}). Such a behavior
has not been seen in any of the other cyclotron line sources. The
second strongest luminosity-dependent cyclotron line variability is
observed in \object{V0332+53}
\citep{Mowlavi2006a,Tsygankov2006a,Nakajima2010a}, in which the
cyclotron lines are so strong that the change can be seen by eye even
in the raw detector spectra. While the boundaries between the
different accretion regimes estimated by \citet{Becker2012a} might
therefore have to be adjusted, the overall principal behavior of the
accretion column discussed by these authors still holds.

With the recently successful launch of \textit{NuSTAR}
\citep{Harrison2010a}, a low background, focusing telescope with
slightly higher resolution than the \textit{RXTE}-HEXTE or the
\textit{Suzaku}-HXD is now available. With its higher resolution and
the much higher expected signal to noise ratio in the spectra, we will
be able, for the first time, to resolve cyclotron lines and study
their luminosity-dependent behavior without the danger of a systematic
influence of the choice of a continuum.

\begin{acknowledgements}
  We thank the referee for his/her insightful comments. We also thank
  the schedulers of \rxte\ and \textit{INTEGRAL} for their role in
  making this campaign possible, and the International Space Science
  Institute in Bern, Switzerland, for their hospitality.  We
  acknowledge funding by the Bundesministerium f\"ur Wirtschaft und
  Technologie under Deutsches Zentrum f\"ur Luft- und Raumfahrt grants
  50OR0808, 50OR0905, and 50OR1113, and by the Deutscher Akademischer
  Austauschdienst. MTW is supported by the US Office of Naval
  Research. IC acknowledges financial support from the French Space
  Agency CNES through CNRS. SMN and JMT acknowledge support from the
  Spanish Ministerio de Ciencia, Tecnolog\'ia e Innovaci\'on (MCINN)
  through grant AYA2010-15431 and the use of the computer facilities
  made available through the grant AIB2010DE-00057. This research is
  in part based on observations with \textit{INTEGRAL}, an ESA project
  with instruments and science data centre funded by ESA member states
  (especially the PI countries: Denmark, France, Germany, Italy,
  Switzerland, Spain), Czech Republic, and Poland, and with the
  participation of Russia and the USA. We thank John E. Davis for the
  development of the \texttt{SLxfig} module, which was used to create
  all figures in the paper.
\end{acknowledgements}

\bibliography{mnemonic,aa_abbrv,references}
\bibliographystyle{aa}

\end{document}